\documentclass[%
 reprint,
 unsortedaddress,
 nofootinbib,
 aps,prd,
 onecolumn,
]{revtex4-2}

\usepackage{babel}
\usepackage{graphicx}
\usepackage{xcolor}
\usepackage{amsmath}
\usepackage{amssymb}
\usepackage{bm}
\usepackage{booktabs}
\usepackage{multirow}
\usepackage{array}
\usepackage[ISO]{diffcoeff}
\usepackage{empheq}
\usepackage{mleftright}
\usepackage{makecell}
\usepackage{siunitx}
\usepackage{hyperref}
\usepackage[capitalise, noabbrev]{cleveref}

\makeatletter
\input{aas_macros.sty}
\let\jnl@style=\relax
\makeatother

\makeatletter
\AtBeginDocument{\immediate\write\@auxout{\string\citation{apsrev42Control}}}

\newlength{\affilindent}
\newcommand*{\affilscriptphantom}{\phantom{\normalfont\textsuperscript{99}}}
\newcommand*{\printaffiliation}[4]{%
    \ifnumequal{#1}{0}{}{%
        \setlength{\affilindent}{\widthof{\affilscriptphantom}}%
        \let\\=\newline%
        \par\noindent%
        \looseness=-100\hangindent=\affilindent%
        \affilscriptphantom\llap{\normalfont\textsuperscript{#1}}%
        \ignorespaces#3%
        \par%
    }%
}

\makeatletter
\appto{\frontmatter@above@affiliation@script}{%
    \global\let\affiliationslist=\@AFF@list%
    \let\@AFF@list=\relax%
    Affiliations are given after the Appendix.%
}{}{}
\newcommand{\printaffiliations}{%
    \begingroup%
    \let\AFF@opr=\printaffiliation%
    \frontmatter@affiliationfont%
    \raggedright%
    \affiliationslist%
    \endgroup%
}

\makeatother

\DeclareSIUnit{\year}{yr}
\DeclareSIUnit{\Gyr}{\giga\yr}
\DeclareSIUnit{\pc}{pc}
\DeclareSIUnit{\kpc}{\kilo\pc}
\DeclareSIUnit{\Mpc}{\mega\pc}
\DeclareSIUnit{\Gpc}{\giga\pc}
\DeclareSIUnit{\Msun}{\text{\ensuremath{M_\odot}}}
\DeclareSIUnit{\hHubble}{\text{\ensuremath{h}}}
\crefname{section}{\S}{\S}
\crefformat{section}{\S#2#1#3}
\crefmultiformat{section}{\S#2#1#3}{ and~\S#2#1#3}{, \S#2#1#3}{ and~\S#2#1#3}
\crefname{appendix}{appendix \S}{appendices \S}
\crefformat{appendix}{appendix \S#2#1#3}
\crefmultiformat{appendix}{appendices \S#2#1#3}{ and~\S#2#1#3}{, \S#2#1#3}{ and~\S#2#1#3}
\crefname{equation}{equation}{equations}
\let\pcref=\cref

\difdef{l}{dn}{style=d^}

\begin{document}

%\preprint{APS/123-QED}

\title{First Detection of the Baryon Acoustic Oscillation (BAO) Feature in the 3-Point Correlation Function of DESI DR1 Luminous Red Galaxies}
% \thanks{A footnote to the article title}

% \author{Farshad Kamalinejad}
% \email{F.Kamalinejad@ufl.edu}
%  \affiliation{Department of Physics, University of Florida, 2001 Museum Rd., Gainesville, FL 32611, USA}
% \author{Zachary Slepian}
% \affiliation{Department of Astronomy, University of Florida, 211 Bryant Space Science Center, Gainesville, FL 32611, USA}
% \author{Alessandro Greco}
% \affiliation{Department of Astronomy, University of Florida, 211 Bryant Space Science Center, Gainesville, FL 32611, USA}
% \author{Alex Krolewski}
% \affiliation{}
% \author{William Ortolá Leonard}
% \affiliation{Department of Physics, University of Florida, 2001 Museum Rd., Gainesville, FL 32611, USA}
% \author{Jessica Chellino}
% \affiliation{Department of Astronomy, University of Florida, 211 Bryant Space Science Center, Gainesville, FL 32611, USA}
% \author{Elena Fernandez Garcia, Simon May, Paco Prada, Anatoly Klypin, Robert Cahn and Jiamin Hou}

\author{Farshad Kamalinejad$^{2, 1}$}
\email{F.Kamalinejad@ufl.edu}
\author{Zachary Slepian$^{2}$}
\author{Alex Krolewski$^{3,4}$}
\author{Alessandro Greco$^{2}$}
\author{William Ortolá Leonard$^{1}$}
\author{Jessica Chellino$^{2}$}
\author{Matthew Reinhard$^{1}$}
\author{Elena Fernández-García$^{5}$}
\author{Francisco Prada$^{5}$}
\author{J.~Aguilar$^{6}$}
\author{S.~Ahlen$^{7}$}
\author{A.~Anand$^{6}$}
\author{C.~Bebek$^{6}$}
\author{D.~Bianchi$^{8,9}$}
\author{D.~Brooks$^{10}$}
\author{T.~Claybaugh$^{6}$}
\author{A.~Cuceu$^{6}$}
\author{K.~S.~Dawson$^{11}$}
\author{A.~de la Macorra$^{12}$}
\author{R.~Demina$^{13}$}
\author{P.~Doel$^{10}$}
\author{J.~Edelstein$^{14,15}$}
\author{J.~E.~Forero-Romero$^{16,17}$}
\author{E.~Gaztañaga$^{18,19,20}$}
\author{S.~Gontcho A Gontcho$^{21}$}
\author{G.~Gutierrez$^{22}$}
\author{H.~K.~Herrera-Alcantar$^{23,24}$}
\author{K.~Honscheid$^{25,26,27}$}
\author{C.~Howlett$^{28}$}
\author{D.~Huterer$^{29,30}$}
\author{M.~Ishak$^{31}$}
\author{R.~Joyce$^{32}$}
\author{S.~Juneau$^{32}$}
\author{D.~Kirkby$^{33}$}
\author{T.~Kisner$^{6}$}
\author{A.~Kremin$^{6}$}
\author{O.~Lahav$^{10}$}
\author{C.~Lamman$^{27}$}
\author{M.~Landriau$^{6}$}
\author{L.~Le~Guillou$^{34}$}
\author{M.~Manera$^{35,36}$}
\author{A.~Meisner$^{32}$}
\author{R.~Miquel$^{37,36}$}
\author{J.~Moustakas$^{38}$}
\author{A.~Muñoz-Gutiérrez$^{12}$}
\author{J.~ A.~Newman$^{39}$}
\author{W.~J.~Percival$^{4,40,3}$}
\author{C.~Poppett$^{6,14,15}$}
\author{I.~P\'erez-R\`afols$^{41}$}
\author{L.~Samushia$^{42,43,44}$}
\author{E.~Sanchez$^{45}$}
\author{D.~Schlegel$^{6}$}
\author{M.~Schubnell$^{29,30}$}
\author{H.~Seo$^{46}$}
\author{J.~Silber$^{6}$}
\author{D.~Sprayberry$^{32}$}
\author{G.~Tarl\'{e}$^{30}$}
\author{B.~A.~Weaver$^{32}$}
\author{C.~Zhao$^{47}$}
\author{H.~Zou$^{48}$}

% \affiliation{$^{1}$Department of Physics, University of Florida, 2001 Museum Rd., Gainesville, FL 32611, USA}
% \affiliation{$^{2}$Department of Astronomy, University of Florida, 211 Bryant Space Science Center, Gainesville, FL 32611, USA}
% \affiliation{$^{3}$Waterloo Centre for Astrophysics, Department of Physics and Astronomy, University of Waterloo, 200 University Avenue West, Waterloo, ON N2L 3G1, Canada}
% \affiliation{$^{4}$Department of Physics and Astronomy, University of Waterloo, Waterloo, ON N2L 3G1, Canada}

% \affiliation{$^{5}$Instituto de Astrof\'{i}sica de Andaluc\'{i}a (CSIC), Glorieta de la Astronom\'{i}a, s/n, E-18008 Granada, Spain \\
% The affiliations of the remaining authors are listed in Appendix E.}

%\author{Anatoly Klypin}
%\affiliation{Department of Astronomy, New Mexico State University, Las Cruces, NM 88003, USA}
%\affiliation{Department of Astronomy, University of Virginia, Charlottesville, VA 22906, USA}

% \author{Robert Cahn}
% \affiliation{Lawrence Berkeley National Laboratory, 1~Cyclotron Road, Berkeley, CA, 94720, USA}

% \author{Jiamin Hou}
% \affiliation{Max-Planck-Institut für Extraterrestrische Physik,
% Postfach~1312, Gießenbachstraße~1,
% 85748 Garching, Germany}
% \affiliation{Institute of Astronomy, University of Cambridge, Madingley Rd., Cambridge CB3 0HA, UK}
% \affiliation{Kavli institute for Cosmology Cambridge, Madingley Rd., Cambridge CB3 0HA, UK}
% \affiliation{Department of Astronomy, University of Florida, 211~Bryant Space Science Center, Gainesville, FL 32611, USA}

%\date{\today}

\begin{abstract} 
We present the first detection of the 3-Point Correlation Function (3PCF) Baryon Acoustic Oscillation (BAO) signal from the DESI Data Release 1 (DR1) sample of Luminous Red Galaxies (LRGs), which contains over 2.1 million galaxies. Our analysis is based on a tree-level redshift-space bispectrum template, which is then transformed to position space using the Fast Fourier Transform on Logarithmic scales (FFTLog)  algorithm. We detect the BAO feature with a significance of approximately $8.1\sigma$ using the \texttt{EZmock} covariance matrix and $8.5\sigma$ using the analytical covariance matrix, for the full LRG redshift range ($0.4<z<1.1$), denoted as the $z_{\rm full}$ sample. We use the \texttt{Abacus altMTL} mocks, the most precise DESI DR1 mock catalogs currently available, to validate our model. We find that our model fits the mocks well, with a small offset of \SI{0.6}{\percent} in the recovered BAO scale, which we treat as a systematic error due to modeling. We measure the angle-averaged distance, $D_{\rm V}(z = 0.68)/r_{\rm d} = \num{15.88 +- 0.27}$ ($1.72\%$ precision) when using the covariance matrix estimated from \texttt{EZmocks} and $D_{\rm V}(z = 0.68)/r_{\rm d} = \num{15.72 +- 0.18}$ ($1.12\%$ precision) when using the analytical Gaussian covariance matrix. Our results show excellent agreement with the DESI DR1 2PCF BAO measurements as well. We also explore several other ways to estimate the error and find between $1.7$--$2.2\%$ precision on the BAO scale from the \texttt{EZmock} covariance matrix and between $1.1$--$1.5\%$ precision from the analytical covariance matrix. This work represents the first detection of the BAO feature in the DESI 3PCF, establishing its ability to probe the expansion history of the Universe with future DESI 3PCF measurements.
\end{abstract}

%\keywords{Suggested keywords}  % Use showkeys class option if keyword display desired

\maketitle

%\tableofcontents

\section{Introduction}
\label{sec:intro}

Baryon Acoustic Oscillations (BAO) \cite{Eisenstein:1997ik} have been frequently used in modern cosmological analyses, including studies of the Cosmic Microwave Background (CMB) and galaxies' 3D clustering in the Large-Scale Structure (LSS) of the Universe. BAO provide a robust method for measuring cosmic distances, enabling stringent tests of cosmological models and constraints on cosmological parameters. The physical origin of BAO lies in the early Universe, prior to recombination, when baryons and photons were tightly coupled in a hot, dense plasma \cite{Ma:1995ey}. Unlike Cold Dark Matter (CDM), which interacts only gravitationally with this plasma, photons and electrons were strongly coupled via Thomson scattering, causing the photon–baryon fluid to behave as a single tightly bound component \cite{Slepian:2014dda}. Quantum fluctuations seeded during inflation led to density and pressure perturbations in this fluid. The resulting photon pressure pushed matter outward from over-dense regions, dragging baryons along and generating acoustic waves in the medium \cite{SE_BAO_2016, Bashinsky:2002vx, Bashinsky:2000uh, Eisenstein:2006nj}. These waves propagated at the sound speed $c_{\mathrm{s}}$, which is set by the baryon-to-photon momentum density ratio:\footnote{The factor of $3/4$ arises because baryons contribute only to the energy density, whereas photons contribute to both the energy density and the pressure.}
\begin{align}
    c_{\mathrm{s}}(z)     = \frac{c}{\sqrt{3 \left(1+\left(3/4\right)\rho_{\mathrm{b}}(z)/\rho_{\gamma}(z)\right)}}
    ,
    \label{eq:sound_speed}
\end{align}
where $c$ is the speed of light, $\rho_{\mathrm{b}}(z)$ is the baryon density and $\rho_{\gamma}(z)$ is the photon density, and both are functions of redshift.
Once the temperature of the Universe dropped below approximately \SI{3000}{\K}, photons decoupled from baryons and began to stream freely. At this point, the sound waves stalled, imprinting a characteristic scale \cite{Eisenstein_2007}---roughly \SI{100}{\per\hHubble\Mpc}---into the distribution of baryons. In detail, the baryon velocity at decoupling has a sharp feature at this scale, known as ``velocity overshoot'' \cite{1980ApJ...236..323P}, and the mode of the density perturbation that starts growing after decoupling inherits the spatial structure of the velocity perturbation \cite{SE_BAO_2016}. The characteristic scale is given by:
\begin{align}
    r_{\mathrm{d}} = \int_{z_{\mathrm{d}}}^{\infty}\frac{c_{\mathrm{s}}(z)}{H(z)} \, \dl z,
    \label{eq:rd}
\end{align}
where $z_{\mathrm{d}}$ is the redshift of the drag epoch ($z_{\mathrm{d}} = {1059.39 \pm 0.46}$ \cite{Planck:2018vyg}) and $H(z)$ is the Hubble parameter at redshift $z$. Since dark matter interacts only gravitationally with the rest of the components, it gradually converges toward the distribution of baryons over time \cite{Eisenstein_2007}. As a result, a spherical shell of excess matter forms around each initial over-density, leaving a relic imprint of the high-redshift acoustic oscillations. This imprint manifests today as the BAO feature in the large-scale distribution of matter.

The BAO are also observed in the 3D distribution of galaxies. As mentioned earlier, the growing mode of the baryon density inherits the spatial structure of the velocity at decoupling, and this leads to an excess of baryons and eventually (due to gravity) dark matter on a spherical shell at the BAO scale. In a matter-dominated Universe, galaxies will eventually form from these excess baryons---initially through the collapse of baryonic matter on small scales, and later through hierarchical mergers. Also, since the scale of this spherical shell is known (which is $147.09 \pm 0.26\; {\rm Mpc}$) \cite{Planck:2018vyg}, we can use the BAO as a standard ruler. A standard ruler is a scale that we know and can use to measure the angular diameter distances in the Universe. 

Let us explain in more detail. Imagine a pair of galaxies separated by a distance of $r_{\mathrm{d}}$ (which can be calculated from \pcref{eq:sound_speed,eq:rd}) at redshift $z$. If the pair of galaxies is aligned along the line of sight we can measure the redshift interval $\Delta z$ (which is small compared to $z$), translating into a distance of $D_{\mathrm{H}}(z) = c/H(z) = r_{\mathrm{d}}/\Delta z$. If the pair of galaxies is oriented perpendicular to the line of sight, we can measure the angle $\Delta \theta$, and therefore determine the transverse comoving distance $D_{\mathrm{M}}(z) = r_{\mathrm{d}}/\Delta \theta$. Therefore, BAO measurements constrain both quantities $D_{\mathrm{H}}(z)/r_{\mathrm{d}}$ and $D_{\mathrm{M}}(z)/r_{\mathrm{d}}$ \cite{DESI:2025zgx}.

% spherical shell of size $r_{\mathrm{d}}$ (which can be calculated from \pcref{eq:sound_speed,eq:rd}) at redshift $z$. If the spherical shell lies perpendicular to the line of sight we can measure the redshift interval $\Delta z$ (which is small compared to $z$), translating into a distance of $D_{\mathrm{H}}(z) = c/H(z) = r_{\mathrm{d}}/\Delta z$. If the shell is perpendicular to the line of sight, we can measure the angle $\Delta \theta$, and therefore determine the transverse comoving distance $D_{\mathrm{M}}(z) = r_{\mathrm{d}}/\Delta \theta$. Therefore, BAO measurements constrain both quantities $D_{\mathrm{H}}(z)/r_{\mathrm{d}}$ and $D_{\mathrm{M}}(z)/r_{\mathrm{d}}$ \cite{DESI:2025zgx}.

The BAO measurement using LSS galaxy statistics is carried out by using the $N$-Point Correlation Functions (NPCF) of galaxies (\cite{2026enap....5..251S} for a review). The NPCF represents the excess probability of finding $N$ galaxies in a particular configuration over the background. In the simplest case, the 2PCF measures the excess probability of finding a pair of galaxies at distance $r$ where the excess probability is evaluated relative to a spatially random distribution \cite{Jeong2015, SE_Full3PCF_BAO, Liu:2017fbp}. The 2PCF decreases as $r$ increases. At $r \simeq \SI{100}{\per\hHubble\Mpc}$, there is a ``bump" in the 2PCF, which is due to the BAO \cite{SE_BAO_2016, peebles_1970, sunyaev_1970, sakh_1966, Ansarinejad:2018xsh, Eisenstein_2005, Hong:2015dqk, Hong:2012gc, Abdullah_2023}. The 2PCF of galaxies depends on the cosmological model and its parameters and, therefore, a full-shape analysis of the 2PCF can constrain both \cite{DESI:2025ejh, DESI:2024hhd, Ishak:2024jhs}.

BAO in the galaxies' 3D clustering were first detected in 2005 by both the 2-degree Field Galaxy Redshift Survey (2dFGRS; using the power spectrum \cite{2dFGRS:2005yhx}) and the Sloan Digital Sky Survey (SDSS; using the 2PCF \cite{Eisenstein_2005}). Subsequently, SDSS~III \cite{Percival:2007yw, SDSS:2009ocz}, the WiggleZ Dark Energy Survey \cite{Blake:2011wn, Blake:2011en, Kazin:2014qga}, and the 6-degree Field Galaxy Survey \cite{Beutler:2011hx, Carter:2018vce} measured the BAO scale up to a few percent precision. The SDSS Baryon Oscillation Spectroscopic Survey (BOSS) \cite{Anderson_2012, BOSS:2013rlg, Alam_2017} and its extension to higher redshifts, eBOSS \cite{eBOSS:2017cqx}, gave percent-level precision on the location of the peak. The most precise measurement so far has been given by the Dark Energy Spectroscopic Instrument (DESI) \cite{DESI2022.KP1.Instr} 2PCF and power spectrum \cite{DESI:2024mwx, DESI:2025zgx, DESI:2025zpo, DESI:2024lzq, Moon:2023jgl, DESI:2025fxa}. DESI is a highly multiplexed spectroscopic survey instrument that will map over 63 million galaxies \cite{DESI2022.KP1.Instr}, exceeding the original forecast of 39 million objects \cite{DESI2016b.Instr}, across 17,000 deg$^2$ of the sky over an eight-year observing program. It is capable of simultaneously obtaining spectra for approximately 5,000 targets across a $\sim$$3^\circ$ field of view \cite{Corrector.Miller.2023, FiberSystem.Poppett.2024}. For further details on the DESI software pipelines and associated data products, we refer the reader to \cite{Spectro.Pipeline.Guy.2023, SurveyOps.Schlafly.2023}. Cosmological analyses are currently in progress and will be further enabled by the forthcoming DR2 data release \citep{DESI.DR2.DR2}.

The next higher-order statistic is the 3PCF. Unlike the 2PCF, the analysis of the 3PCF has been limited in the past \cite{Pugno:2024uxc, Euclid:2025diy, Farina:2024jfs, SE_Full3PCF_BAO, Moresco_2021, Pearson_2018, fry_1984_Npoint, fry_cluster_galaxy_3pcf, Gazta_aga_2009, Moresco:2020quj}. This limitation is primarily due to the complexity of modeling, measurements, and systematics of the 3PCF \cite{NovellMasot:2025fju}. The 3PCF measurements of galaxies start with modeling the redshift-space bispectrum (the 3PCF in Fourier space) \cite{Scoccimarro_2015, Burger:2025meh}. Numerous studies have investigated the bispectrum of galaxies, showing the potential of the bispectrum in constraining Primordial Non-Gaussianity (PNG) \cite{Desjacques:2010nn, Baldauf_2012, Sefusatti_2007, Baldauf_2011, Shirasaki_2021, Sefusatti:2010ee}, tightening parameter constraints \cite{kamalinejad2020non, bispectrumneutrinolensing, D_Amico_2024, child1, de_Belsunce_2019}, and breaking degeneracies between parameters \cite{Yankelevich_2019, hahn2020constraining, Hahn_2021}.

Additionally, the 3PCF provides a methodologically independent measurement of the BAO scale \cite{Pearson:2017wtw, SE_Full3PCF_BAO, Behera:2023uat, 2018arXiv181112396C, Gaztanaga:2008sq}. As we will see, the full-shape analysis of the galaxy NPCF requires an understanding of galaxy biasing \cite{Desjacques:2018}. Moreover, the DESI survey must be corrected for other sources of complexities and systematics, such as edge effects and fiber collisions \cite{Bianchi:2024fnl, DESI:altMTL, Silber:2022mmp, DESI:2020jeu}. It is crucial to assess whether these systematics impact the BAO measurements derived from the 2PCF. Therefore, the analysis of the 3PCF can independently test the robustness of the treatment of these systematics.

In this paper, we first present an accurate model of the 3PCF based on the redshift-space bispectrum model of \cite{Scoccimarro_1999}. We then use the Fast Fourier Transform on Logarithmic scales (FFTLog) algorithm to compute the 3PCF. The previous 3PCF model applied to the SDSS BOSS CMASS sample \cite{SE_Full3PCF_BAO} relied on an analytical formulation. Our approach can be modified in the future to include higher-order correction terms arising from mode coupling in Fourier space, which is an advantage over purely analytical modeling \cite{SE_3PCF_model}. We then use our model on both the DESI DR1 data and the \texttt{Abacus altMTL} mocks \cite{DESI:altMTL} to measure the BAO scale.

The structure of this paper is as follows: In \cref{sec:Modeling3PCF}, we describe our 3PCF model which includes three free galaxy bias parameters: $b_1$, $b_2$, and $b_{\mathcal{G}_2}$. In \cref{sec:2dFFTlog}, we explain how to transform the bispectrum template into the 3PCF of galaxies. This transformation is performed using a 2D FFTLog, which is both fast and accurate. In \cref{sec:DESIsample}, we describe the sample used—namely, the Luminous Red Galaxy (LRG) sample from DESI DR1--the covariance matrices, the measurement procedure and the range of radial scales we used. In \cref{sec:BAO detection Sig.} we will discuss the detection significance of the BAO signal in the 3PCF of DESI DR1. Finally, in \cref{sec:comparison}, we compare our measurements of the BAO scale with the \texttt{Abacus altMTL} mocks~\cite{DESI:altMTL, abacus, Maksimova:2021ynf}, a suite of high-resolution $N$-body simulations specifically designed for DESI analyses, and demonstrate the agreement between the data and the mocks, and obtain the modeling systematics. We then add this systematic offset (which is \SI{0.6}{\percent}) to the statistical error obtained from the measurement of the BAO scale in the data.

\section{Modeling the 3PCF}
\label{sec:Modeling3PCF}

The quantum fluctuations in the inflationary era seeded the initial conditions of structure formation. These primordial perturbations, which were nearly Gaussian and almost scale-invariant, evolved under the influence of gravity during the radiation- and matter-dominated eras, eventually forming the large-scale structures we observe today. At early times—such as during recombination—these fluctuations had small amplitudes relative to the background and are therefore considered linear. However, by the low-redshift Universe ($z < 1$), gravitational evolution has amplified them to the point where they are no longer small, and the density and velocity fields have become non-Gaussian. In the linear regime, the power spectrum $P_{\mathrm{lin}}(k)$ encapsulates all the statistical information because Wick's theorem (or in our context, Isserlis' theorem) holds \cite{isserlis}: odd-point correlation functions vanish, and even-point ones can be derived from the two-point function. In contrast, non-Gaussianities—whether originating from primordial physics or generated by nonlinear gravitational evolution—lead to nonzero higher-order statistics that carry rich additional information about the underlying cosmology. In what follows we briefly review PT and introduce the bispectrum monopole with respect to the line of sight template in redshift space.

In the linear regime, Fourier modes evolve independently. However, as gravitational non-linearities grow, modes become coupled. This coupling is described by mode-coupling kernels, \textit{i.e.} the perturbation theory kernels $F_n$ and $G_n$, used to express the density field ($\delta$) and the divergence of the velocity field ($\theta$) at all orders as \cite{Bernardeau_2002}:\footnote{Unless stated otherwise, all parameters refer to their present-time values; time dependence is made explicit when relevant.}
\begin{align}
    \delta(\mathbf{k}, \tau) &= \sum_{n=1}^{\infty} D^n(\tau) \int F_n(\mathbf{q}_1,\cdots,\mathbf{q}_n) \, \delta_{\mathrm{D}}^{[3]}\mleft(\mathbf{k}-\sum_{j = 1}^n \mathbf{q}_j\mright) \prod_{i=1}^{n} \delta^{(1)}(\mathbf{q}_i) \frac{\dl.dn.[3]{\mathbf{q}_i}}{(2\pi)^3}
    ,
    \label{Eq:delta_n}
    \\
    \theta(\mathbf{k}, \tau) &= -\mathcal{H}(\tau)\sum_{n=1}^{\infty} D^n(\tau) \int G_n(\mathbf{q}_1,\cdots,\mathbf{q}_n) \, \delta_{\mathrm{D}}^{[3]}\mleft(\mathbf{k}-\sum_{j = 1}^n \mathbf{q}_j\mright) \prod_{i=1}^{n} \delta^{(1)}(\mathbf{q}_i) \frac{\dl.dn.[3]{\mathbf{q}_i}}{(2\pi)^3}
    .
    \nonumber
%    \label{Eq:theta_n}
\end{align}
Here, $D(\tau)$ is the linear growth factor as a function of conformal time,\footnote{$\displaystyle\tau \equiv \int_0^{\tau} \dl a/[a^2 H(a)].$} which scales as $D(\tau) \propto a(\tau)$ in a matter-dominated Universe. $\mathcal{H}(\tau)$ is the Hubble parameter as a function of the conformal time ($\mathcal{H}(\tau) = a(\tau)H(\tau)$). $n$ denotes the order of non-linear correction and $\delta^{(1)}$ is the linear density field. The three-dimensional Dirac $\delta$-distribution $\delta_{\mathrm{D}}^{[3]}$ enforces momentum conservation. The PT kernels $F_n$ and $G_n$, computed via recursion relations \cite{Bernardeau_2002}, describe the nonlinear evolution. Since $F_1 = G_1 = 1$, the second-order kernels are \cite{Bernardeau_2002}:
\begin{align}
    F_2(\mathbf{q}_1, \mathbf{q}_2) = \frac{5}{7}+\frac{1}{2}\frac{\mathbf{q}_1\cdot \mathbf{q}_2}{q_1 q_2}\left(\frac{q_1}{q_2}+\frac{q_2}{q_1}\right) + \frac{2}{7}\left(\frac{\mathbf{q}_1\cdot \mathbf{q}_2}{q_1 q_2}\right)^2,
    \\
    G_2(\mathbf{q}_1, \mathbf{q}_2) = \frac{3}{7}+\frac{1}{2}\frac{\mathbf{q}_1\cdot \mathbf{q}_2}{q_1 q_2}\left(\frac{q_1}{q_2}+\frac{q_2}{q_1}\right) + \frac{4}{7}\left(\frac{\mathbf{q}_1\cdot \mathbf{q}_2}{q_1 q_2}\right)^2.
    \nonumber
\end{align}

In Fourier space, the bispectrum is the ensemble average (``$\langle \rangle$" in the equation below) of three density fields:
\begin{align}
    B(\mathbf{k}_1, \mathbf{k}_2, \mathbf{k}_3)\;(2\pi)^3\delta_{\mathrm{D}}^{[3]}(\mathbf{k}_1+ \mathbf{k}_2+ \mathbf{k}_3) = \langle \delta(\mathbf{k}_1) \delta(\mathbf{k}_2) \delta(\mathbf{k}_3) \rangle.
\end{align}
Here, $B(\mathbf{k}_1, \mathbf{k}_2, \mathbf{k}_3)$ is the bispectrum, with the Dirac $\delta$-distribution ensuring that $\mathbf{k}_1$, $\mathbf{k}_2$, and $\mathbf{k}_3$ form a closed triangle, stemming from translation invariance. In the linear regime, this ensemble average vanishes. However, by expanding the density field to second order as in \cref{Eq:delta_n} and contracting the fields—yielding products of two power spectra—we obtain the real-space bispectrum (i.\,e.\ without redshift-space effects) as \cite{Bernardeau_2002, Scoccimarro_2015}:
\begin{align}
    B(\mathbf{k}_1, \mathbf{k}_2, \mathbf{k}_3) = 2F_2(\mathbf{k}_1, \mathbf{k}_2) P_{\mathrm{lin}}(k_1) P_{\mathrm{lin}}(k_2) + \text{cyc.}
\end{align}
Here, ``cyc." means summing over all cyclic permutations of the triangle configuration of the wave-vectors, $(\mathbf{k}_1, \mathbf{k}_2, \mathbf{k}_3)$, \textit{i.\,e.}, summing over $(\mathbf{k}_1, \mathbf{k}_2, \mathbf{k}_3)$, $(\mathbf{k}_2, \mathbf{k}_3, \mathbf{k}_1)$, and $(\mathbf{k}_3, \mathbf{k}_1, \mathbf{k}_2)$. This enforces symmetry under relabeling.

In practice, when analyzing a galaxy redshift survey, we need to account for additional complexities. First, we do not observe the underlying (dark) matter field directly, but rather the galaxies, which are biased tracers of the density field. Second, observing galaxies in redshift space introduces distortions along the line of sight and mixes the density and velocity perturbations. As a result, the tree-level bispectrum is influenced not only by the $F_2$ kernel but also by the $G_2$ kernel. Additionally, the bispectrum in redshift space depends on two angular variables with respect to the line of sight. The first is the angle between the line of sight ($\hat{z}$) and $\hat{k}_1$, denoted by $\mu$, and the second is the azimuthal angle of $\mathbf{k}_2$ around $\mathbf{k}_1$, denoted by $\phi$. Consequently, the redshift-space bispectrum depends on five variables in total. We expand it in terms of spherical harmonics, $\mathrm{Y}_{\ell m}(\mu, \phi)$, where $\ell$ and $m$ denote the degree and order of the spherical harmonics, respectively, following \cite{Scoccimarro_1999}:
\begin{align}
    B(\mathbf{k}_1, \mathbf{k}_2, \mathbf{k}_3, \mu, \phi)
    = \sum_{\ell, m} B_{(\ell,m)}(\mathbf{k}_1, \mathbf{k}_2, \mathbf{k}_3)\,
    \mathrm{Y}_{\ell m}(\mu, \phi),
\end{align}
and obtain the bispectrum monopole, $B_{(0,0)}$, by averaging over the angular dependence. Calculating the redshift-space bispectrum monopole (with respect to the line of sight), we find \cite{Scoccimarro_1999}:
\begin{align}
    B_{(0, 0)}(\mathbf{k}_1, \mathbf{k}_2, \mathbf{k}_3) &=\Bigg[F_2(\mathbf{k}_1, \mathbf{k}_2) \mathcal{D}_{\mathrm{sq1}}(\beta, \mu_{12})+ G_2(\mathbf{k}_1, \mathbf{k}_2) \mathcal{D}_{\mathrm{sq2}}(\beta, k_1, k_2, \mu_{12})\nonumber\\&\qquad+\mathcal{D}_{\mathrm{FOG}}(\beta, k_1, k_2, \mu_{12})+ \mathcal{D}_{\mathrm{NLB}}(\beta, \gamma, \mu_{12}) + \mathcal{D}_{\mathcal{G}_2}(\beta, \gamma', \mu_{12})\Bigg]b_1^4 P_{\mathrm{lin}}(k_1) P_{\mathrm{lin}}(k_2) + \text{cyc.}
    \label{eq:redshift_space_B}
\end{align}
Here, $\beta \equiv f / b_1$ with $f \equiv \difs{\ln\mleft(D(a)\mright)}{a}$ denoting the logarithmic growth rate, $b_1$ is the linear galaxy bias and $\mu_{12} \equiv \hat{k}_1 \cdot \hat{k}_2$ is the angle between $\mathbf{k}_1$ and $\mathbf{k}_2$. The terms $\mathcal{D}_{\mathrm{sq1}}$ and $\mathcal{D}_{\mathrm{sq2}}$ describe large-scale infall, which squashes over-densities along the line of sight due to gravity. $\mathcal{D}_{\mathrm{FOG}}$ stands for Finger of God (FOG), but in this model this is not actually due to thermal velocities as one might have thought; rather, this term just captures deterministic velocities from the redshift-space PT kernels. $\mathcal{D}_{\mathrm{NLB}}$ captures non-linear biasing with $\gamma \equiv b_2 / b_1$, while $\mathcal{D}_{\mathcal{G}_2}$ represents the tidal tensor contribution with $\gamma' \equiv b_{\mathcal{G}_2} / b_1$. All terms are defined in \cite{Scoccimarro_1999}, except for $\mathcal{D}_{\mathcal{G}_2}$:\footnote{Please refer to \cite{Scoccimarro_1999} for these functions; they are not repeated here.}
\begin{align}
    \mathcal{D}_{\mathcal{G}_2}(\beta, \gamma', \mu_{12}) = \gamma'(\mu_{12}^2-1)D_{\mathrm{sq1}}(\beta, \mu_{12}).
\end{align}

The bispectrum monopole depends only on three variables: $k_1$, $k_2$, and $\mu_{12}$ (or equivalently $k_3$), related by the law of cosines:
$k_3^2 = k_1^2 + k_2^2 + 2 k_1 k_2 \mu_{12}$.
This allows us to expand the bispectrum monopole $B_{(0,0)}$ in terms of Legendre polynomials $\mathcal{L}_{\ell}(\mu_{12})$. The bispectrum inner multipoles (projected onto Legendre polynomials in $\mu_{12}$, i.\,e.\ the cosine of the internal angle between $\mathbf{k}_1$ and $\mathbf{k}_2$) are then defined as:
\begin{align}
    B_{\ell}(k_1, k_2) = \frac{2\ell+1}{2}\int_{-1}^{1}B_{(0,0)}(\mathbf{k}_1, \mathbf{k}_2, \mathbf{k}_3)\,\mathcal{L}_{\ell}(\mu_{12}) \dl\mu_{12}.
\end{align}
Now, we note that the bispectrum inner multipole $B_{\ell}(k_1, k_2)$ is a two-dimensional function depending only on $k_1$ and $k_2$. 
The bispectrum inner multipoles are crucial for computing the 3PCF of galaxies in configuration space. This is because the full 3PCF is the inverse Fourier transform of the redshift-space bispectrum monopole. However, this inverse transform simplifies by integrating over the Dirac $\delta$-distribution after expanding the exponential as a plane wave. This projection onto the angle between $\mathbf{k}_1$ and $\mathbf{k}_2$, denoted by $\mu_{12}$, yields the 3PCF inner multipoles $\zeta_\ell$:
\begin{equation}
    \zeta_{\ell}(r_1, r_2) = (-1)^{\ell} \int \frac{k_1^2 k_2^2 \dl k_1 \dl k_2}{(2\pi^2)^2} B_{\ell}(k_1, k_2) \, j_{\ell}(k_1 r_1) \, j_{\ell}(k_2 r_2),
    \label{eq:3pcfmultipole}
\end{equation}
where the indices $\ell$ correspond to both the bispectrum multipole and the spherical Bessel functions $j_\ell$.

These 3PCF inner multipoles relate to the full 3PCF $\zeta$ through a Legendre expansion:
\begin{align}
    \zeta(r_1, r_2, \hat{r}_1\cdot \hat{r}_2) = \sum_{\ell=0}^{\infty} \zeta_{\ell}(r_1, r_2) \, \mathcal{L}_{\ell}(\hat{r}_1\cdot \hat{r}_2).
\end{align}
In practice, we truncate this series since only a finite number of multipoles are accessible. Typically, multipoles up to $\ell=4$ suffice to capture most relevant information in the galaxy 3PCF \cite{SE_3PCF_model, SE_3pt_alg}. The 3PCF multipoles $\zeta_{\ell}(r_1, r_2)$ can be computed either analytically as double Bessel transforms—as first shown in \cite{SE_3PCF_model}—or numerically using the FFTLog method \cite{simonovic2018cosmological, mcewen2016fast}. In the following, we describe how to calculate $\zeta_{\ell}(r_1, r_2)$.

\subsection{2D FFTLog 3PCF Model}
\label{sec:2dFFTlog}

Here we evaluate the integral in \cref{eq:3pcfmultipole} using the FFTLog method \cite{simonovic2018cosmological, mcewen2016fast, umeh, Fang:2020vhc, guidi_3pcf}. We begin by expanding the bispectrum multipoles as a sum over power laws, sampled at $N$ points on a logarithmic grid in $k_1$ and $k_2$, ranging from $k_{\mathrm{min}}$ to $k_{\mathrm{max}}$ (i.\,e.\ $k_{j} = k_{\mathrm{min}} \exp\mleft[j \ln(k_{\mathrm{max}} / k_{\mathrm{min}}) / (N - 1)\mright]$, $j \in [0, N - 1]$), as \cite{Fang:2020vhc}:
\begin{align}
    B_{\ell}(k_1, k_2) = \frac{1}{N^2} \! \sum_{m, n = -N/2}^{N/2} \!\! c_{mn} k_{\mathrm{min}}^{-i\eta_m}k_{\mathrm{min}}^{-i\eta_n} k_1^{\nu+i\eta_m}k_2^{\nu+i\eta_n} ,
    \qquad
    \eta_m = \frac{2\pi m}{\ln\mleft(k_{\mathrm{max}}/k_{\mathrm{min}}\mright)}
    \label{eq:bk},
\end{align}
where $c_{mn}$ are the coefficients that encode the dependence on the cosmology. ${\nu}$ is a real number called the \textit{bias parameter} (which should not be mistaken for the linear galaxy bias $b_1$), which is chosen such that the integrals we wish to evaluate (Fourier transforms used to obtain the 3PCF) are not divergent. We will discuss our choice of $\nu$ in the following paragraphs. The $c_{mn}$ coefficients are given by a discrete Fourier transform as:
\begin{equation}
    c_{mn} = \sum_{p, q= 0}^{N-1} \frac{B_{\ell}(k_{1, p}, k_{2, q})}{k_{1, p}^{\nu} \, k_{2, q}^{\nu}} \exp\mleft(\frac{-2\pi i \left(mp+nq\right)}{N}\mright)
    .
\end{equation}
These coefficients obey a symmetry relation such that $c^*_{m,n} = c_{-m, -n}$. Substituting \cref{eq:bk} into \cref{eq:3pcfmultipole} and taking the integral analytically, we can obtain the corresponding 3PCF multipole as a sum over $m$ and $n$. 

In reality, we need to bin the 3PCF template which means that we are averaging the 3PCF multipole on grids of $r_1$ and $r_2$. Fortunately, the binning can also be performed analytically. Doing so, we obtain the binned 3PCF as:
\begin{align}
    \bar{\zeta}_{\ell}(r_i, r_j) = \frac{(-1)^{\ell}}{\mathcal{N}} 
    \sum_{m,n} \frac{c_{mn}}{8\pi^3}\,
    g\mleft(\ell+\tfrac{1}{2}, z_m - 1/2\mright)\,
    g\mleft(\ell+\tfrac{1}{2}, z_n - 1/2\mright)\,
    \left(\frac{r_i^{2-z_m}}{2-z_m}\Bigg|_{r_{i, {\mathrm{min}}}}^{r_{i, {\mathrm{max}}}}\right)
    \left(\frac{r_j^{2-z_n}}{2-z_n}\Bigg|_{r_{j, {\mathrm{min}}}}^{r_{j, {\mathrm{max}}}}\right)
    \label{eq:binned3pcf}.
\end{align}
$r_i$ and $r_j$ are the central values of the $i^{\text{th}}$ and $j^{\text{th}}$ radial bin. $\mathcal{N}$ is the normalization factor and is $\mathcal{N} = \mleft(r_{i, {\mathrm{max}}}^3 - r_{i, {\mathrm{min}}}^3\mright) \mleft(r_{j, {\mathrm{max}}}^3 - r_{j, {\mathrm{min}}}^3\mright) / \, 9$, and $z_{m(n)} = 2 + \nu + i\eta_{m(n)}$.
$g({\ell}, z)$ is defined as:
\begin{align}
   g(\ell, z) \equiv 2^z \frac{\Gamma\mleft[\left(\ell+z+1\right)/2\mright]}{\Gamma\mleft[\left(\ell-z+1\right)/2)\mright]}.
\end{align}
The integral in \cref{eq:3pcfmultipole} can be evaluated analytically using the expansion in \cref{eq:bk} only if $-3 - \ell < \nu < -1$. Otherwise, the integrals will be divergent \cite{Fang:2020vhc}. These conditions define an allowed range for the bias parameter $\nu $.

Let us now summarize the settings used in our analysis. In the bispectrum template, we adopt $\smash{k_{\mathrm{min}} = \SI{e-5}{\hHubble\per\Mpc}}$ and $\smash{k_{\mathrm{max}} = \SI{5}{\hHubble\per\Mpc}}$, with 500 sampling points for both $k_1$ and $k_2$. To suppress ringing effects and reduce the contribution from the high-$k$ tail of the power spectrum—where linear theory is no longer valid—we apply a Gaussian damping factor of the form $\exp\mleft[-(k/k_{\mathrm{high}})^2\mright]$, with $k_{\mathrm{high}} = \SI{1}{\hHubble\per\Mpc}$. We have tested that any damping scale in the range $\SI{1}{\hHubble\per\Mpc} \leq k_{\mathrm{high}} $ works. Additionally, we fix the bias parameter to $\nu = -2.1$, which ensures convergence of the algorithm for all relevant multipoles. We note that we have performed convergence tests for different resolutions and scale cuts and the settings reported here are efficient enough for this analysis.

\subsection{Wiggle vs. No-Wiggle}
\label{sec:W-vs-NoW}

As explained earlier, the BAO signature in the power spectrum appears as a series of oscillations (or ``wiggles") spanning the range $\SI{0.01}{\hHubble\per\Mpc} \lesssim k \lesssim \SI{0.2}{\hHubble\per\Mpc}$ for a typical cosmology. In the 2PCF of galaxies, these wiggles transform into a single, localized bump at a scale of roughly \SI{100}{\hHubble\per\Mpc}. Following the methodology applied to the BOSS CMASS sample \cite{SE_Full3PCF_BAO}, the BAO detection significance is obtained by fitting a 3PCF template to the data and evaluating the corresponding $\chi^2$. This procedure is repeated after removing the wiggles from the power spectrum, resulting in the so-called ``no-wiggle" power spectrum, denoted by $P_{\mathrm{nw}}(k)$.

There are several methods for removing the BAO wiggles from the power spectrum. The simplest approach is to use the Eisenstein and Hu no-wiggle transfer function $T(k)$ \cite{Eisenstein:1997ik}, since the matter power spectrum can be expressed as $P(k) \propto k^{n_{\mathrm{s}}} T(k)^2$ ($n_{\mathrm{s}}$ is the primordial tilt). This method was employed by \cite{SE_Full3PCF_BAO}. In this work, however, we adopt a different technique introduced in \cite{class_pt} and originally developed by \cite{Hamann_2010} for incorporating IR (Infra-Red) resummation \cite{blas2016time, Blas_2016}. We begin by sampling $\ln\mleft(k\,P(k)\mright)$ on a logarithmic grid in $k$ with $2^{16}$ points. We then apply the Fast Sine Transform (FST) to this array and decompose the result into even and odd harmonics \footnote{Alternatively, one could take the Fourier transform of the power spectrum, localize the BAO bump in configuration space, remove it, and then transform back to obtain the smooth power spectrum. However, this approach is computationally more complex, as it requires an FFTLog decomposition for efficient evaluation of the Fourier integrals.}. The BAO feature manifests as a localized bump in both the even and odd harmonics, which can then be removed through interpolation if desired. After subtracting the bump, we apply the inverse FST to reconstruct a smooth, no-wiggle power spectrum.

\begin{figure}
    \centering
    \includegraphics[width=\textwidth]{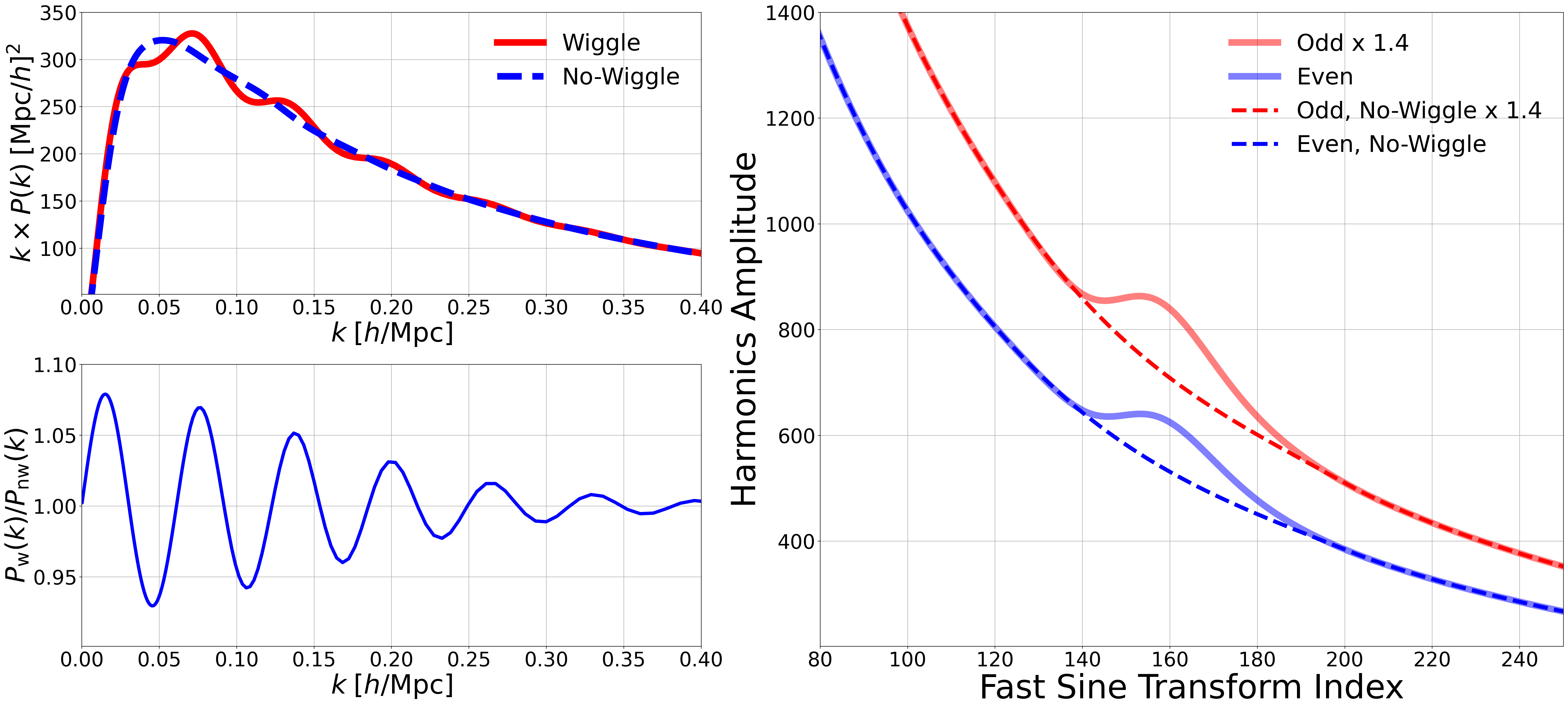}

    \caption{\textit{Left:} The top panel shows the wiggle power spectrum (red) and the no-wiggle power spectrum (blue), while the bottom panel displays their ratio. \textit{Right:} We present the odd and even harmonics of the Fast Sine Transform (FST) of the power spectrum \cite{class_pt, Hamann_2010}. Even harmonics are shown in blue (solid for the wiggle spectrum and dashed for the no-wiggle spectrum), and odd harmonics are shown in red using the same line style convention. The horizontal axis corresponds to the FST index, not to physical radial distances. We observe that the BAO bump spans FST indices from approximately 140 to 200 for our fiducial cosmology. We have verified that this range is roughly consistent across other similar cosmologies. To separate them on the plot from the even harmonics, we have multiplied the odd harmonics by 1.4. Overall, these wiggle and no-wiggle power spectrum models are used to estimate the BAO detection significance. In this plot we used $z = 0.68$, as it is the effective redshift of the $z_{\mathrm{full}}$ LRG sample (as further described in the \S\ref{sec:DESIsample}).}
    \label{Fig:wigglevsnow}
\end{figure}

\Cref{Fig:wigglevsnow} shows the result of this procedure. In the right column, the even harmonics of the FST are shown as the blue curves: the solid line corresponds to the wiggle power spectrum, while the dashed line represents the same spectrum with the BAO bump removed. The odd harmonics are displayed in red, with the solid line showing the wiggle spectrum and the dashed line the spectrum with the bump removed. For better visualization, the odd harmonics have been multiplied by a factor of 1.4 so that both sets can be shown clearly on the same plot. In the left column, the top panel shows the wiggle and no-wiggle power spectra in red and blue respectively, while the bottom panel displays their ratio.

Throughout this work, we fix the cosmology to the \textit{Planck} 2018 baseline \cite{Planck:2018vyg}, which is also adopted by the \texttt{AbacusSummit} simulation suite \cite{Maksimova:2021ynf, abacus}, a set of high-resolution $N$-body simulations designed for DESI. The cosmological parameters used are: $\omega_{\mathrm{cdm}} \equiv \Omega_{\mathrm{cdm}} h^2 = 0.1200$ (dark matter density), $\omega_{\mathrm{b}} \equiv \Omega_{\mathrm{b}} h^2 = 0.02237$ (baryon density), $A_{\mathrm{s}} = \num{2.0830e-9}$ (amplitude of primordial fluctuations), $h = 0.6736$ (Hubble constant in units of \SI{100}{\km\per\s\per\Mpc}), and $n_{\mathrm{s}} = 0.9649$ (scalar spectral index). We also include three massive neutrino states with a total mass of $\sum m_{\nu} = \SI{0.06}{\eV}$. 

Although recent DESI BAO and full-shape power spectrum analyses favor cosmological models with redshift-dependent dark energy over a cosmological constant \cite{DESI:2024hhd, DESI:2024uvr, DESI:2024jis}, we adopt a fixed $\Lambda$CDM\footnote{$\Lambda$ stands for the cosmological constant and CDM for Cold Dark Matter.} model for this work and use the Alcock–Paczynski (AP) factor $\alpha$ \cite{Alcock:1979mp} to account for the potential difference between the true cosmology and the fiducial one.

\begin{figure}
    \centering
    \includegraphics[width=0.5\textwidth]{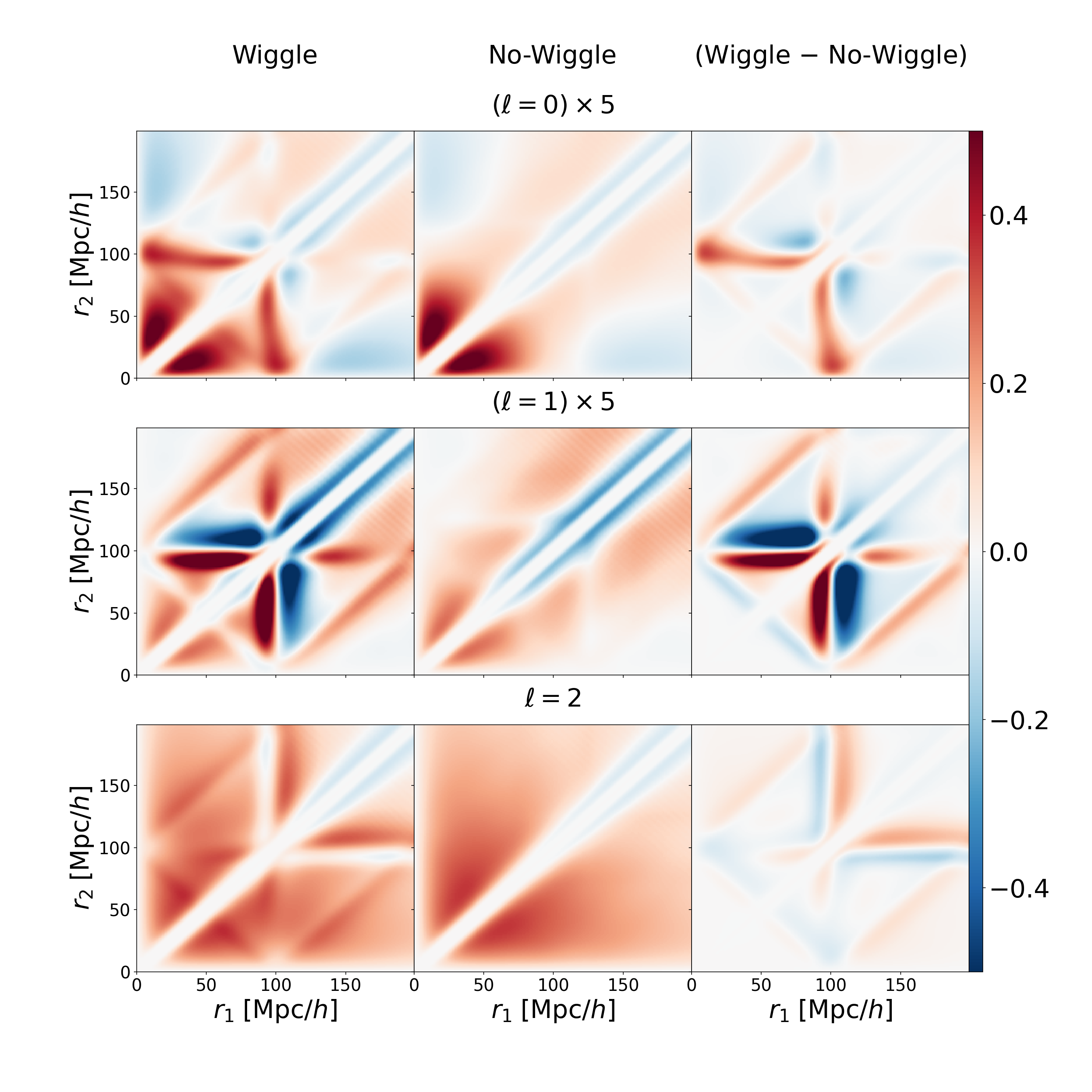}%
    \hfill%
    \includegraphics[width=0.5\textwidth]{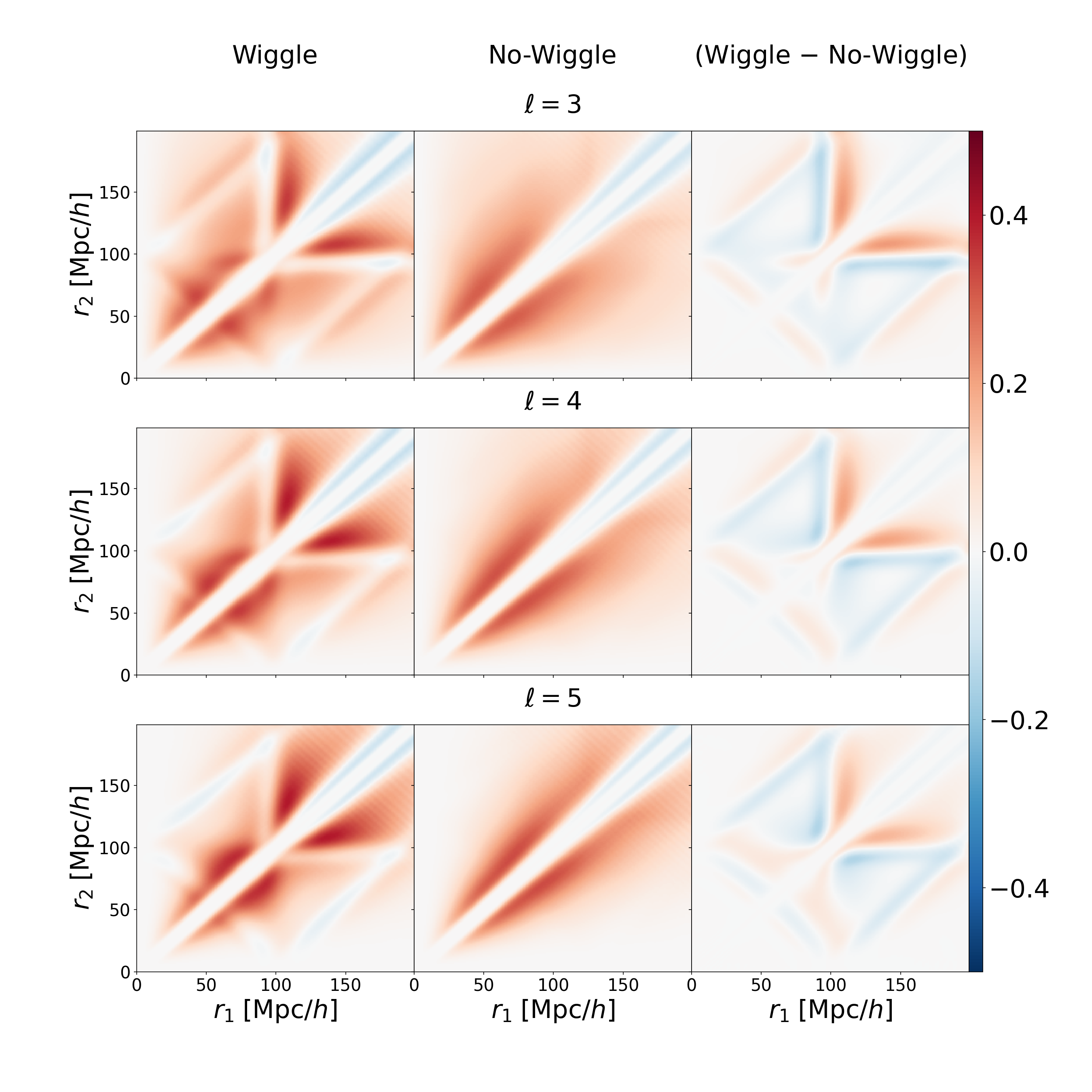}

    \caption{We compare the 3PCF multipoles up to $\ell = 5$ obtained from the wiggle power spectrum (\textit{left columns} in each group of panels), which contains the BAO signal, and from the no-wiggle power spectrum (\textit{middle columns}), which has the BAO feature removed. The \textit{right columns} show the difference between the two. Each panel is also multiplied by the damping factor $\mathcal{D}_{\mathrm{visualization}}$, as defined in \cref{eq:ddamp}. Also, the $\ell = 0$ and $\ell = 1$ panels are multiplied by five for better visualization. As seen in the comparison, the BAO feature appears clearly around the characteristic scale $r_{\mathrm{d}} = \SI{100}{\per\hHubble\Mpc}$. In addition to this main scale, we observe enhanced signal when $|r_1 \pm r_2| = r_{\mathrm{d}}$, corresponding to configurations sensitive to the BAO geometry. The same features around $|r_1 \pm r_2| = r_{\mathrm{d}}$ are also seen in other works such as \cite{SE_Full3PCF_BAO, Chellino:2025dbi, Leonard:2024xvy}. The presence of these features in the 3PCF will enable us to estimate the BAO detection significance using the DESI DR1 data in the following sections. This plot is at $z = 0.68$, the effective redshift of the $z_{\mathrm{full}}$ LRG sample (as explained further in \S\ref{sec:DESIsample}).}
    \label{fig:3PCF_Panel}
\end{figure}

Now, with the wiggle and no-wiggle power spectra at hand, we use \cref{eq:binned3pcf} to compute the corresponding wiggle and no-wiggle 3PCF templates. We note that the power spectrum enters the bispectrum via \cref{eq:redshift_space_B}. As a first step, we examine the unbinned 3PCF to visualize the impact of removing the BAO feature. \Cref{fig:3PCF_Panel} shows the 3PCF multipoles computed using the wiggle power spectrum, $\zeta_{\mathrm{w}}(r_1, r_2)$ (all panels on the left), the no-wiggle power spectrum, $\zeta_{\mathrm{nw}}(r_1, r_2)$ (middle panels), and their difference, $\zeta_{\mathrm{w}}(r_1, r_2) - \zeta_{\mathrm{nw}}(r_1, r_2)$ (right panels). 

For better visualization, all plots have been multiplied by the damping factor
\begin{align}
    \mathcal{D}_{\mathrm{visualization}}(r_1, r_2) = \frac{r_1^2 r_2^2}{b_1^3 \left(\SI{10}{\per\hHubble\Mpc}\right)^4} \exp\mleft(-\frac{\left(\SI{12}{\per\hHubble\Mpc}\right)^2}{(r_1 - r_2)^2}\mright),
    \label{eq:ddamp}
\end{align}
following the approach of \cite{SE_Full3PCF_BAO}. We weight all of the plots by $\mathcal{D}_{\mathrm{visualization}}$ since the 3PCF scales as $1/(r_1^2 r_2^2)$ on large scales. Additionally, since we do not include configurations with $r_1 = r_2$ in our analysis (since these configurations include squeezed triangles with a vanishing third side), we apply an exponential factor to suppress them. Finally, we divide by $b_1^3$, as the amplitude of the 3PCF scales with the cube of the linear bias, $b_1^3$.

We observe that the BAO feature is clearly present across all 3PCF multipoles. In particular, a pronounced signal appears around the BAO scale, $r_{\mathrm{d}} \sim \SI{100}{\per\hHubble\Mpc}$. Additionally we see a visible BAO imprint on the 3PCF when $r_1 + r_2 = r_{\mathrm{d}}$ and when $|r_1 - r_2| = r_{\mathrm{d}}$, reflecting the geometric nature of the BAO signature (the same feature is also seen in figure~6 of \cite{SE_Full3PCF_BAO} and also in \cite{child2, SE_3pt_alg}). The $\ell = 0$ and $\ell = 1$ panels are scaled by a factor of five to enhance visibility. These results are obtained assuming the fiducial cosmology described earlier, at a redshift of $z = 0.68$ (which is the effective redshift of the $z_{\mathrm{full}}$ sample as we further detail later), with fiducial galaxy bias parameters $b_1 = 1.75$, $b_2 = 1$, and $b_{\mathcal{G}_2} = -0.2$.

\section{3PCF of DESI Data Release I}
\label{sec:DESIsample}

The DESI DR1 galaxy catalog \cite{DESI:2024aax} consists of four different tracers: BGS (Bright Galaxy Sample), ELGs (Emission Line Galaxies), LRGs (Luminous Red Galaxies), and QSOs (Quasars). The BGS sample covers the redshift range $0.1 < z < 0.4$ and contains \num{300 043} galaxies. The ELG sample spans $0.8 < z < 1.6$ with \num{2 432 072} galaxies, while the LRG sample includes \num{2 138 627} galaxies in the redshift range $0.4 < z < 1.1$. Finally, the QSO sample contains \num{856 831} quasars over the redshift range $0.8 < z < 2.1$. For more information about the DESI samples, target selection, and validation, we refer to \cite{DESI:2022gle, DESI:2016fyo, Lan:2022bdh, DESI:2024mwx, DESI:2024jis}.

In this work, we focus on the 3PCF measurement and modeling of the LRGs. These galaxies are preferred for our analysis because they have large linear bias parameter, have been well-studied in past work (BOSS) \cite{SE_Full3PCF_BAO} and are less susceptible to redshift and imaging systematics because they are bright and have a very high redshift success rate. The DESI DR1 LRG sample is divided into three non-overlapping redshift bins: LRG1 ($0.4 < z < 0.6$), LRG2 ($0.6 < z < 0.8$), and LRG3 ($0.8 < z < 1.1$). The number densities of the LRG1 and LRG2 samples are relatively uniform, with $\bar{n} \sim \SI{5e-4}{\hHubble\cubed\per\Mpc\cubed}$ \cite{DESI:2023ytc, DESI:2024uvr}. In contrast, the number density in the LRG3 sample drops significantly to about \SI{1e-4}{\hHubble\cubed\per\Mpc\cubed} at $z=1.1$, yielding an overall average of $\bar{n} \sim \SI{3.5e-4}{\hHubble\cubed\per\Mpc\cubed}$.

To obtain a larger cumulative volume, we combine the LRG1, LRG2, and LRG3 samples, which have a mean redshift of $\bar{z} = 0.68$, and refer to the combined set as the $z_{\rm full}$ sample. This sample was previously used by \cite{slepian2025measurement} in their 4PCF analysis. In that work, \cite{slepian2025measurement} also combined the LRG1 and LRG2 samples, referring to the result as the $z_{\rm cut}$ sample. Following \cite{Krolewski:2024paz}, \cite{slepian2025measurement} further divided the $z_{\rm cut}$ and $z_{\rm full}$ samples into smaller sub-divisions of the sky, referred to as ``regions'' (for the $z_{\rm cut}$ sample) and ``patches'' (for the $z_{\rm full}$ sample). This division is necessary for the 4PCF analysis~\cite{Krolewski:2024paz, hou2025study, slepian2025measurement} but not for the 3PCF analysis. Therefore, in this paper, we only consider the largest volume which is the entire $z_{\rm full}$ sample (spanning $0.4 < z < 1.1$) without dividing it into smaller patches.

\subsection{Measurement \& Covariance Matrix}
\label{subsec:Measurement}

The measurement of the 3PCF isotropic basis coefficients for the DESI DR1 dataset is performed using \texttt{CADENZA}, a GPU-accelerated version of the \texttt{ENCORE} code~\cite{encore}. The algorithm implemented in \texttt{CADENZA} is based on the method developed in~\cite{SE_3pt_alg, encore}, which we briefly review here. Since the 3PCF involves triplets of galaxies forming triangle configurations, one might expect the computational cost to scale as $\mathcal{O}(N^3)$, where $N$ is the number of galaxies. However, by exploiting rotational symmetries, along with the spherical harmonic addition theorem, the computational complexity is reduced to $\mathcal{O}(N^2)$ which is the same as for the 2PCF. This results in a substantial gain in computational efficiency. The algorithm proceeds by expanding the radially-binned galaxy density field around each galaxy using spherical harmonics on each bin, and then combining these expansions to compute the 3PCF isotropic basis coefficients, $\bar{\zeta}_{\ell}(r_1, r_2)$. This approach naturally accounts for binning and edge correction with minimal additional computational cost (section 3.3 of \cite{encore}).

We measure the 3PCF isotropic basis coefficients, $\bar{\zeta}_{\ell}(r_1, r_2)$, using \texttt{CADENZA} up to $\ell = 4$ (with $\ell = 5$ computed to enable edge correction of the $\ell = 4$ mode). We exclude the $\ell = 5$ isotropic basis coefficients because proper edge correction at this order would require the computation of the $\ell = 6$ multipole as well. As a result, our data vector $\mathbf{D}$ consists of the binned 3PCF isotropic basis coefficients: $\bar{\zeta}_{0}(r_1, r_2)$, $\bar{\zeta}_{1}(r_1, r_2)$, $\bar{\zeta}_{2}(r_1, r_2)$, $\bar{\zeta}_{3}(r_1, r_2)$, and $\bar{\zeta}_{4}(r_1, r_2)$. We use 18 radial bins, which yield 105 unique $(r_1, r_2)$ combinations with $r_1 \neq r_2$ for each multipole. Therefore, the total length of the data vector is $5 \times 105 = 525$. Since we are fitting only for the three galaxy bias parameters (as will be discussed later) and the BAO scale ($\alpha$), the number of degrees of freedom (DOF) in the fit is $525  - 4 = 521$.

\subsubsection{Range of Scales}

The range of scales considered in this analysis is defined as follows. Since we use a tree-level bispectrum template from PT, we do not expect the model to be valid on small scales. Therefore, we exclude any configurations where $|r_1 - r_2| < \SI{20}{\per\hHubble\Mpc}$. We also require both $r_1$ and $r_2$ be larger than \SI{20}{\per\hHubble\Mpc}. This choice results in 105 valid radial configurations (As stated in the previous paragraph), covering the range from $r_{\mathrm{min}} = \SI{23.9}{\per\hHubble\Mpc}$ to $r_{\mathrm{max}} = \SI{153.1}{\per\hHubble\Mpc}$, with bin width $\Delta r = \SI{7.8}{\per\hHubble\Mpc}$.

\subsubsection{Covariance Matrix}

We fit our theoretical template to the data using the covariance matrix estimated from 1000 \texttt{EZmock} (Effective Zeldovich approximation) simulation realizations~\cite{EZmock}. These mocks are first generated and then evolved using the Zeldovich approximation~\cite{Zeldovich:1969sb}, which accounts for the non-linear growth of matter. Galaxy summary statistics are then measured on these mocks and used to estimate the covariance matrix. Since these mocks are not full $N$-body simulations and are based on analytical models, they are very inexpensive to create, relative to an $N$-body simulation suite. The covariances are constructed separately for each patch of the NGC and SGC, taking into account the specific survey geometry and fiber collision effects in each patch. 

There have been two separate methods to mirror the DESI fiber assignment on the mock catalogs, called alternative Merged Target Ledger (\texttt{altMTL}) and Fast Fiber Assign (FFA); both are discussed in more detail in \cite{DESI:2024aax}. Due to the computational expense of \texttt{altMTL}, only FFA, which is much faster, was applied to all of the \texttt{Ezmocks}, which in this work, we use to estimate the covariance matrix.

It has been suggested to rescale the \texttt{EZmock} covariance matrices by a tracer- and redshift-bin–dependent constant, calibrated through comparison with semi-analytical covariance matrices produced by \texttt{RascalC}~\cite{Philcox_2019}. This correction is motivated by the fact that \texttt{EZmocks} constructed with the FFA prescription do not accurately capture the true uncertainties in the pair counts. Such calibrations have not yet been performed for the \texttt{EZmock} realizations in the 3PCF. Such a rescaling would simply increase the covariance matrix, and so if one applied it, it would slightly reduce the detection significance \cite{DESI:2024aax}.%
\footnote{We note that the best-fit $\alpha$ and galaxy bias parameters are unaffected by an overall rescaling of the covariance matrix.} Therefore, the analysis presented in this paper may be regarded as slightly optimistic, up to an overall rescaling of the covariance matrix.

We also note that since the covariance is obtained from a finite number of realizations, while the covariance itself is an unbiased estimator of the true covariance matrix, its inverse is not. Therefore, as suggested by \cite{hartlap}, a correction must be applied to the inverse of the covariance, formally known as the Hartlap factor: $\alpha_{\mathrm{Hartlap}} = (N_{\mathrm{obs}} - N_{\mathrm{bins}} - 2)/(N_{\mathrm{obs}} - 1)$, where $N_{\mathrm{obs}}$ is the number of mocks and $N_{\mathrm{bins}}$ is the size of the data vector (number of bins). In our case, $\alpha_{\mathrm{Hartlap}} = 0.47$, which is applied to the inverse of the covariance matrix.
%\ak{You also need to account for the Percival factor, especially if your Hartlap factor is so large! See pages 39 and 40 of 2411.12021. Will's paper, 190 in 2411.12021, discusses all of this in more detail, and also shows that the likelihood becomes a $t$ distribution. This distinction is irrelevant in more usual applications where the Hartlap factor is tiny, but could matter here.}

An additional effect that must be taken into account when the Hartlap factor is much smaller than unity is the Percival factor. The Percival factor is needed since the covariance matrix itself is noisy and this leads to underestimation of the errorbars in the posterior distribution. The Percival factor applied to the covariance matrix is:
\begin{align}
    \alpha_{\rm Percival} = \frac{1+B(N_{\rm bins} - N_{\rm params})}{1+A+B(N_{\rm params} - 1)},
\end{align}
where 
\begin{align}
    A = \frac{2}{(N_{\rm obs} - N_{\rm bins} - 1)(N_{\rm obs} - N_{\rm bins} - 4)}, \;B = \frac{N_{\rm obs} - N_{\rm bins} - 2}{(N_{\rm obs} - N_{\rm bins} - 1)(N_{\rm obs} - N_{\rm bins} - 4)}.
\end{align}
In this work, $\alpha_{\rm Percival} = 2.10$ which will be applied to the covariance matrix estimated from the 1000 \texttt{EZmock} simulations. It is worth noting that the Percival factor is required when assessing the parameter uncertainties obtained from the posterior, not when measuring the detection significance. Therefore, in our work, we do not apply the Percival factor to the covariance matrix when measuring the detection significance. However, we do apply it when measuring the BAO scale and its uncertainty.

Another approach we can take is to use the analytical 
Gaussian covariance matrix~\cite{SE_3pt_alg, Chellino:2025dbi}, calibrated against the covariance matrix obtained from the \texttt{EZmock}s. In this approach, we do not need to apply the Hartlap or Percival factors to the analytical covariance matrix itself, since it is not estimated from a finite number of \texttt{EZmock} realizations. We fit the analytical covariance matrix template (Eq.~(4.1) of~\cite{Chellino:2025dbi} in isotropic basis or Eq.~(65) of~\cite{SE_3pt_alg} in the Legendre basis) by maximizing the log-likelihood function with respect to the mean number density $\bar{n}$ and the effective volume $V_{\rm eff}$~\cite{Hou:2021ncj, slepian2025measurement}:
\begin{align}
    -\log{\mathcal{L}(\bar{n}, V_{\rm eff})} \propto 
    {\rm Tr}\!\left[C^{-1}_{\rm analytical}(\bar{n}, V_{\rm eff})\, C_{\texttt{EZmock}}\right] 
    - \log\!\left(\det C^{-1}_{\rm analytical}(\bar{n}, V_{\rm eff})\right).
\end{align}
Note that we do not invert $C_{\texttt{EZmock}}$. Fitting the analytical covariance template to $C_{\texttt{EZmock}}$ yields $\bar{n} = (1.48\pm 0.01) \times 10^{-4}\,[h/{\rm Mpc}]^3$ and $V_{\rm eff} = 0.84\pm 0.01\,[{\rm Gpc}/h]^3$ for the $z_{\rm full}$ sample for the NGC, and $\bar{n} = (0.96 \pm 0.01)\times 10^{-4}\,[h/{\rm Mpc}]^3$ and $V_{\rm eff} = 0.61\pm 0.01\,[{\rm Gpc}/h]^3$ for the SGC.

\section{BAO Detection Significance and Distance Measurement}
\label{sec:BAO detection Sig.}

One of our primary concerns is the accurate estimation of uncertainties in the 3PCF. In our analysis, we use both the \texttt{Ezmocks} covariance matrix and the calibrated analytical covariance matrix. We also employ four different methods to estimate the error bars on the BAO scale, which we refer to as the standard method, the bootstrapping method, the mock scatter method and the marginalization method. These methods provide different and independent estimates of the BAO scale error bars. We adopt the standard method as the official result of this work, while the other methods are discussed for comparison. We discuss the bootstrapping, mock scatter and the marginalization methods in Appendices 
\S\ref{sec:Bootstrap} ,\S\ref{sec:MocksMethods} and \S\ref{sec:MargMethod}, respectively.

Our standard method can be summarized as follows. We fit the wiggle and no-wiggle galaxy 3PCF templates, each with three free parameters ($b_1$, $b_2$, and $b_{\mathcal{G}_2}$), to the DESI DR1 data to obtain the best-fit galaxy bias parameters. Since the cosmological parameters are not explicitly varied in the fit, we re-scale the $k$-modes by an AP factor, $\alpha$, to effectively vary the combination of the cosmological parameters. $\alpha$ is chosen to lie within the range $0.7 \leq \alpha \leq 1.3$ \cite{Alcock:1979mp, Padmanabhan_2008, Carter_2020, SE_Full3PCF_BAO}. We note that a single scaling factor $\alpha$ suffices because the 3PCF used here is averaged over triangle orientations and so no longer depends explicitly on the line of sight.

Once the best-fit bias parameters are determined for each model (wiggle and no-wiggle) and for each value of $\alpha$ in the range  $0.7 \leq \alpha \leq 1.3$, we compute the goodness of fit using the $\chi^2$ statistic:
\begin{align}
    \chi^2 = (\mathbf{D} - \mathbf{M})^{\mathrm{T}}\,\mathbf{C}^{-1}(\mathbf{D} - \mathbf{M}),
\end{align}
where $\mathbf{D}$ and $\mathbf{M}$ are the data and model vectors, respectively, and $\mathbf{C}^{-1}$ is the inverse of the covariance matrix.

After computing $\chi^2(\alpha)$, we identify the minimum $\chi^2$ value for the wiggle model and denote the corresponding AP parameter as $\alpha_{\mathrm{min}}$. We then evaluate the difference in chi-squared between the wiggle and no-wiggle models at this best-fit value:
\begin{align}
    \Delta \chi^2 = \chi^2_{\mathrm{nw}}(\alpha_{\mathrm{min}}) - \chi^2_{\mathrm{w}}(\alpha_{\mathrm{min}}).
\end{align}
%\ak{Why not re-optimize $\alpha$} separately for wiggle and no wiggle
The BAO detection significance, in terms of the standard deviation of the Gaussian used for the likelihood, is given by $\sqrt{\Delta \chi^2}$. we combine the detection significances from NGC and SGC by adding the individual significances in quadrature to obtain the total significance.

% We repeat this procedure for each patch and region listed in \cref{tab:Regions}. Since the patches are statistically independent (i.\,e., they do not overlap), we may combine the detection significances by adding the individual $\sigma$ values in quadrature to obtain the total significance. 

% An important caveat arises when the value of $\Delta \chi^2$ for a patch is negative. Statistically, this is allowed (and, as we will show, does occur). A negative $\Delta \chi^2$ indicates that the data in that patch prefers the no-wiggle model over the wiggle model. In such cases, when summing the $\sigma$ values in quadrature, the corresponding significance must be included with a negative sign.

Based on the $\chi^2$s we found for our samples, we can estimate the AP parameter $\alpha$. To do so, we first interpolate the total $\chi^2$ of the wiggle model, then find the range that corresponds to $\chi^2 = \chi^2_{\mathrm{min}}+1$. This will give us the $1\sigma$ error bar on $\alpha$. We denote this statistical error bar as $\sigma_{\mathrm{stat.}}$. Once the precision on $\alpha$ is obtained, we define the angle-averaged distance as the mean of distances measured along and across the line of sight \cite{Eisenstein_2005, Anderson_2012}:
\begin{align}
    D_{\mathrm{V}}(z) \equiv\left[z \,D_{\mathrm{M}}(z)^2\,D_{\mathrm{H}}(z)\right]^{1/3} = \left[c\,z \,(1+z)^2 D_{\mathrm{A}}^2(z)/H(z)\right]^{1/3}.
    \label{eq:Dv}
\end{align}
$D_{\mathrm{M}}(z)$ and $D_{\mathrm{H}}(z)$ are defined in this work's introduction (\cref{sec:intro}) and $D_{\mathrm{A}}$ is the angular diameter distance. The ratio $D_{\mathrm{V}}/r_{\mathrm{d}}$ is expressed in terms of the AP parameter, $\alpha$ as:
\begin{align}
    D_{\mathrm{V}}(z)/r_{\mathrm{d}} = \alpha\times \left(D_{\mathrm{V}}(z) /r_{\mathrm{d}}\right)^{\mathrm{fid}}.
    \label{Eq:dv/rd}
\end{align}
where $\left(D_{\mathrm{V}}(z) /r_{\mathrm{d}}\right)^{\mathrm{fid}}$ is the fiducial value of the $D_{\mathrm{V}}$ divided by $r_{\mathrm{d}}$. We measure the angle-averaged distance normalized by the comoving sound horizon at the drag epoch $r_{\mathrm{d}}$, and its $1\sigma$ error bar, obtained from the measurement of $\alpha$. The cosmology we assumed yields $(D_{\mathrm{V}}(z = 0.6) /r_{\mathrm{d}})^{\mathrm{fid}} = 14.59$ for the $z_{\mathrm{cut}}$ sample and $(D_{\mathrm{V}}(z = 0.68) /r_{\mathrm{d}})^{\mathrm{fid}} = 16.02$ for the $z_{\mathrm{full}}$ sample.

In addition to the statistical uncertainties in the estimation of $\alpha$, there can also be systematic uncertainties due to our modeling. Ideally, these systematic uncertainties would be zero, but there is no way to measure them directly from the data. Therefore, we need to compare our model against mock catalogs. We will discuss the mocks used later in \cref{sec:comparison}. For now, let us explain the methodology.

We fit our wiggle 3PCF model to the mocks and obtain the best-fit $\alpha_{\mathrm{min}}$ for each mock. If the model perfectly described the mocks, the mean value of $\alpha$ would be exactly $\alpha_{\mathrm{min}} = 1$, with no offset. However, this is not always the case, and we do not recover a mean value of $\alpha$ exactly equal to 1 across the mocks. We therefore interpret this offset as the \textit{systematic} uncertainty of our modeling. Consequently, the BAO scale can be expressed as:
\begin{align}
    \alpha = \alpha_{\mathrm{min}} \pm (\sigma_{\mathrm{sys.}}) \pm (\sigma_{\mathrm{stat.}}) ,
\end{align}
where $\alpha_{\mathrm{min}}$ and $\sigma_{\mathrm{stat.}}$ are obtained from the data (as explained above \pcref{eq:Dv}), and $\sigma_{\mathrm{sys.}}$ is estimated from the mocks. The total uncertainty on the BAO scale is then obtained from adding $\sigma_{\mathrm{stat.}}$ and $\sigma_{\mathrm{sys.}}$ in quadrature.

\begin{figure}[htbp]
    \centering
    % First plot
    \includegraphics[width=0.99\textwidth]{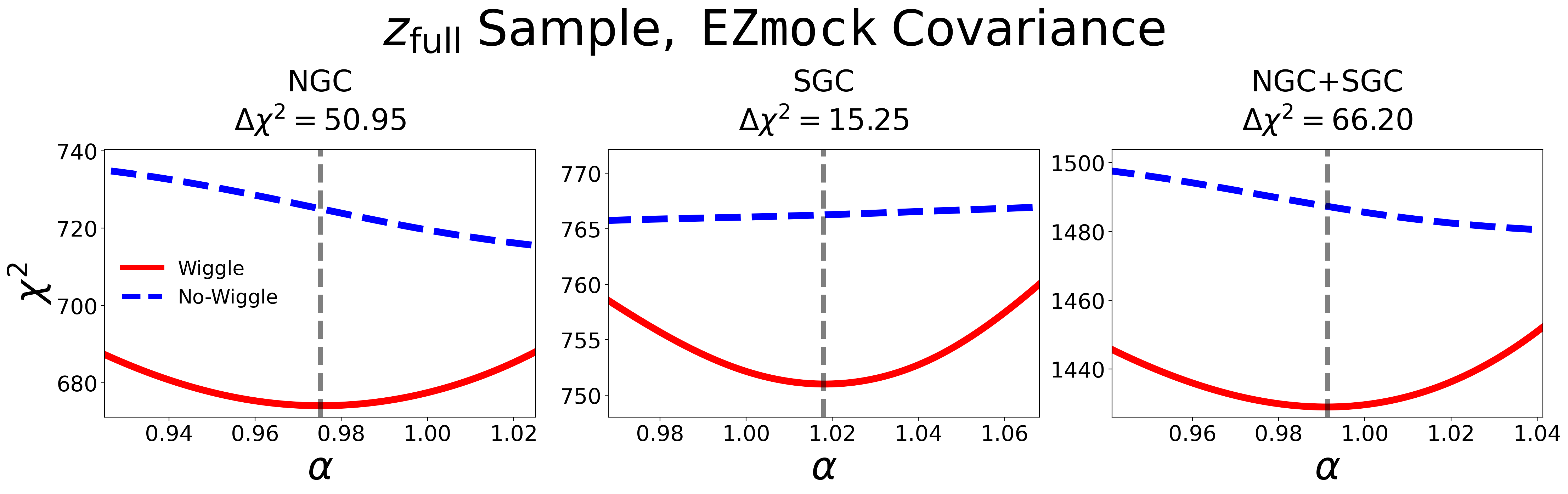}
    
    % Second plot below
    \includegraphics[width=0.99\textwidth]{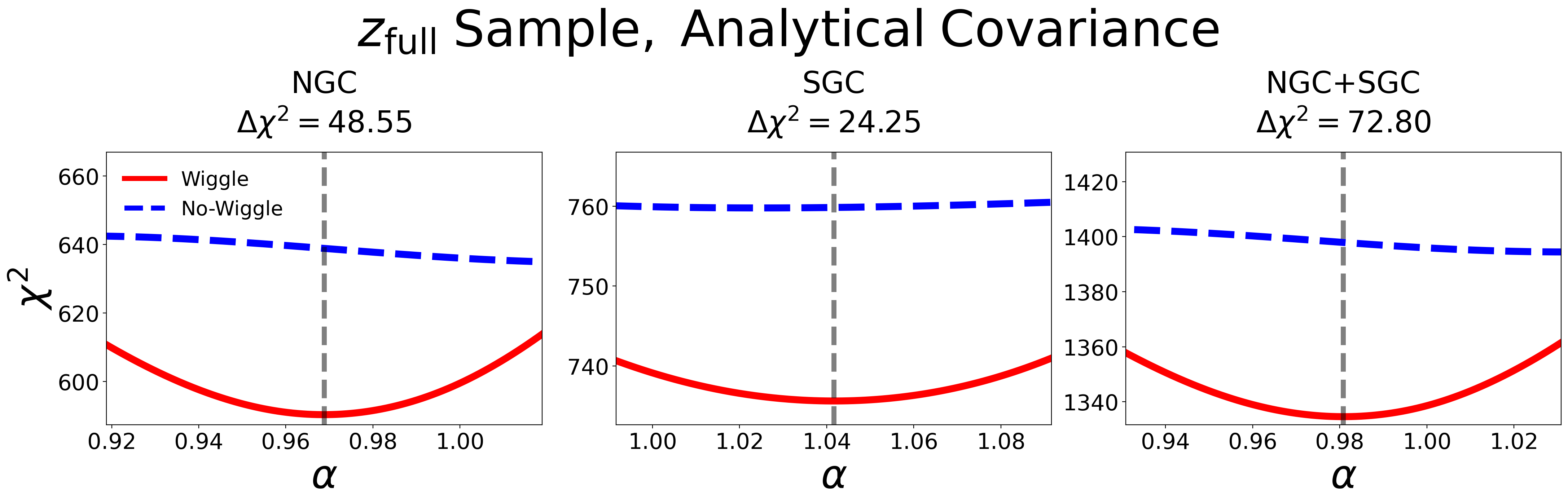}

    \caption{The $\chi^2$ values from fitting the 3PCF template to the DESI DR1 data using the wiggle (solid red) and no-wiggle (dashed blue) models are shown for the NGC and SGC of the $z_{\mathrm{full}}$ sample. The difference between the $\chi^2$ of the wiggle and no-wiggle when using the analytical covariance matrix yields a $\sim$$7.0\sigma$ detection significance from the NGC and $\sim$$5.0\sigma$ detection significance from the SGC. When combining the detection significances in quadrature, we obtain a total significance of approximately $8.5\sigma$. Using the covariance matrix estimated from \texttt{EZmock}s, the total detection significance is $8.1\sigma$. The parameter $\alpha$ is included in the analysis to account for the difference between the fiducial cosmology and the true cosmology. The vertical gray dashed lines indicate the location of $\alpha_{\mathrm{min}}$, which corresponds to the minimum of $\chi^2_{\mathrm{w}}$ for the wiggle model. The last panel is obtained by summing the $\chi^2$ values from NGC and SGC for the wiggle and no-wiggle models separately.}
    \label{Fig:zfulltotal}
\end{figure}

\Cref{Fig:zfulltotal} shows the $\chi^2$ of the wiggle and no-wiggle models to the NGC and SGC using the \texttt{Ezmock} covariance matrix (top panel) and the analytical covariance matrix (bottom panel). Using the \texttt{Ezmock} covariance matrix, we can detect the BAO signal at $8.1\sigma$, while using the analytical covariance matrix results in an $8.5\sigma$ detection, which is in perfect agreement to the \texttt{Ezmock} covariance matrix case. In \S\ref{sec:Rescale}, we discuss how rescaling the analytical covariance matrix to make $\chi^2$/dof exactly equal to 1 affects the BAO measurements.
% \begin{figure}
%     \centering
%     \includegraphics[width=0.99\textwidth]{Images/chi2zful_total.png}

%     \caption{The $\chi^2$ values from fitting the 3PCF template to the DESI DR1 data using the wiggle (solid red) and no-wiggle (dashed blue) models are shown the NGC and SGC of the $z_{\mathrm{full}}$ sample. The difference between the $\chi^2$ of the wiggle and no-wiggle yields a $7.1\sigma$ detection significance from the NGC and $3.91\sigma$ detection significance from the SGC. When combining the detection significances in quadrature, we obtain a total significance of approximately $8.1\sigma$. The parameter $\alpha$ is included in the analysis to account for the difference between the fiducial cosmology and the true cosmology. The vertical gray dashed lines indicate the location of $\alpha_{\mathrm{min}}$, which corresponds to the minimum of $\chi^2_{\mathrm{w}}$ for the wiggle model. The last panel (Total) is obtained by summing the $\chi^2$ values from NGC and SGC for the wiggle and no-wiggle models separately.}
%     \label{Fig:zfulltotal}
% \end{figure}

\section{Systematic Offset \& Comparison with Mocks}
\label{sec:comparison}

Let us now compare our results with the same analysis that we outlined in this paper, but on the \texttt{Abacus-2} \texttt{altMTL} mocks \cite{abacus, Maksimova:2021ynf}. The simulation suite contains 25 realizations, each with a \SI{2}{\per\hHubble\Gpc} box size. These are the most realistic mocks designed for DESI DR1, incorporating the \textsc{fiberassign} code, which has been shown to reproduce the DESI fiber assignment on DESI data \cite{DESI:altMTL}. We validate the analysis of the 3PCF presented earlier on these realizations.

The procedure is the same as for the data, as we explained before. We measure the 3PCF data on each simulation box, then fit the 3PCF wiggle and no-wiggle models to them. We measure the galaxy biases as well as the AP parameter $\alpha$ for each realization. As before, we fix the cosmology to the \textit{Planck} 2018 \cite{Planck:2018vyg}, which is the underlying model for the mocks as well. The mocks we use are at the redshift of our $z_{\mathrm{full}}$ sample, which is at $z_{\mathrm{eff}} = 0.68$, and therefore, we can compare the measurements of the $z_{\mathrm{full}}$ sample with the mocks. Hence, we can find the $\chi^2$ values of the fits. Measuring the $\chi^2$ and the $\alpha$ parameter across all realizations enables us to obtain the distribution of these quantities and, therefore, find the Probability Distribution Function (PDF) of them. Once the PDF is obtained, we can compare the $\chi^2$ and $\alpha$ with the data.

\begin{figure}[h]
    \centering
    \includegraphics[width=0.98\textwidth]{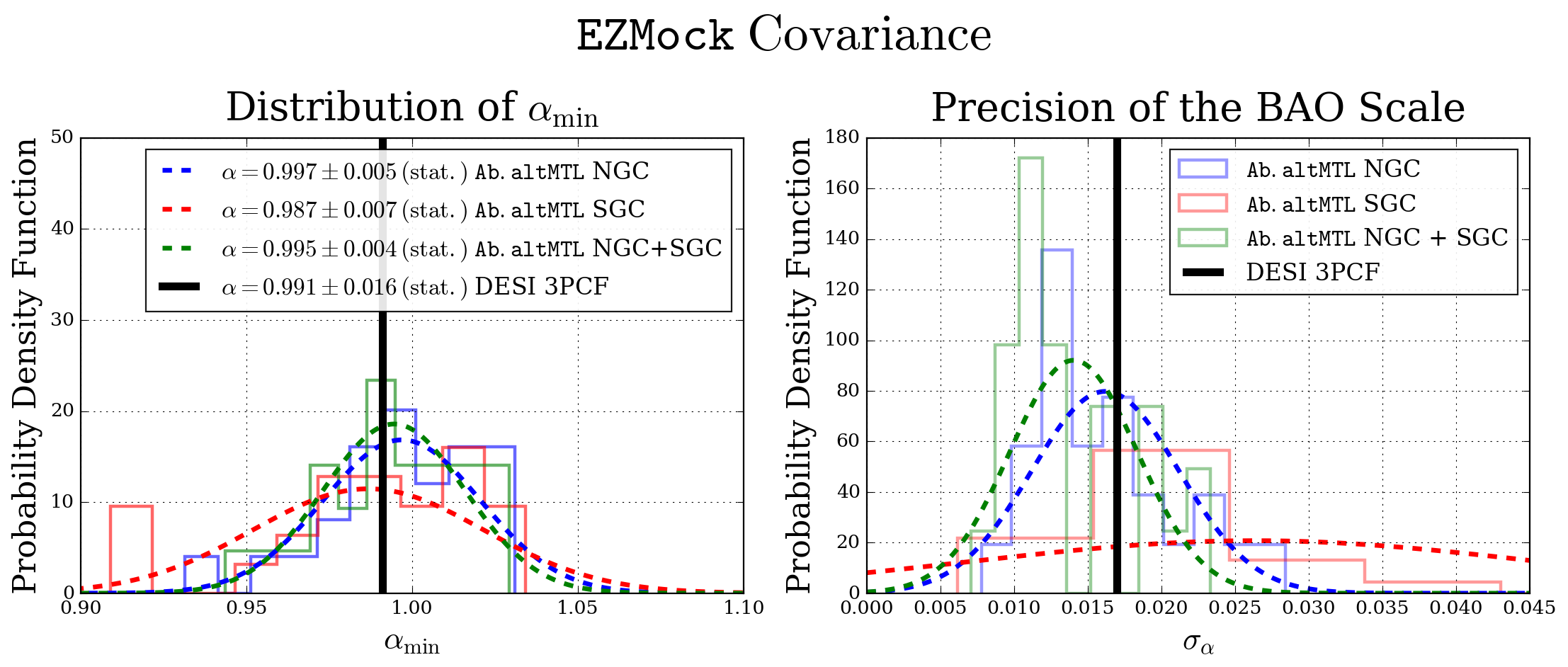}
    \vspace{0.3cm}
    \includegraphics[width=0.98\textwidth]{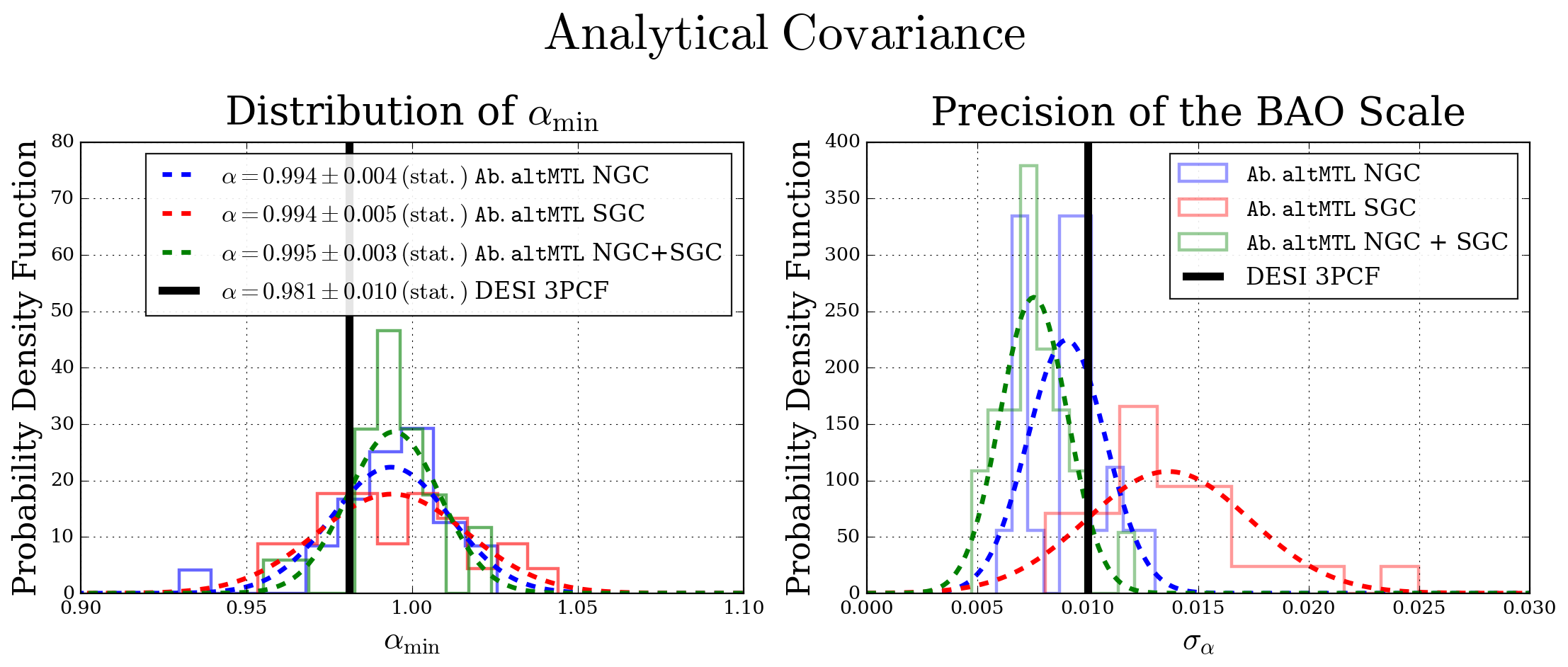}
    \caption{\textit{Left:} 
    Probability Distribution Function (PDF) of $\alpha_{\mathrm{min}}$, which is the value of $\alpha$ at which the $\chi^2$ of the wiggle model is minimized. The blue, red, and green histograms correspond to the distribution of $\alpha_{\mathrm{min}}$ obtained respectively from the NGC, SGC, and NGC+SGC of the \texttt{Abacus altMTL} mocks. The solid black vertical line represents the value from the data. The legends denote the means of the histograms as well as the uncertainty on the means, which is the standard deviation of the PDF divided by the square root of the number of mocks. This panel shows that the estimate of $\alpha_{\mathrm{min}}$ from the data is completely consistent with the mocks. We also note that the (stat.) errors given in the legends are the uncertainty of the mean, not the standard deviation of the Gaussian. \textit{Right:} PDF of the precision on the BAO scale, denoted as $\Delta \alpha$. We measure $\Delta \alpha$ for each mock, similarly to the data, by minimizing $\chi^2_{\mathrm{w}}$ and then finding the $1\sigma$ width of $\alpha$. As in the left panel, the blue, red, and green histograms correspond repectively to the NGC, SGC, and NGC+SGC of the \texttt{Abacus altMTL} mocks, while the solid black vertical line is for the data. We note that, since the NGC constrains the BAO scale with higher precision, it drives the precision of the combined measurement. The data appear to give slightly worse precision on the BAO scale than the NGC+SGC of the mocks (green). We see that the statistical error on $\alpha$ is completely consistent with the mocks.}
    \label{fig:alpha_EZmock}
\end{figure}

Let us present the comparison. \Cref{fig:alpha_EZmock} shows the comparison of the $\alpha_{\rm min}$ (the BAO scale itself) and $\sigma_\alpha$ (the precision of the BAO scale) we obtained from the mocks vs.\ the measurement we presented in \cref{sec:BAO detection Sig.} by using the \texttt{EZmock} covariance and the analytical covariance, respectively. The left panel of both plots shows the distribution of $\alpha_{\mathrm{min}}$, which is the $\alpha$ where $\chi^2_{\mathrm{w}}$ is minimized ($\chi^2$ of the wiggle model). Blue, red, and green represent the distribution of $\alpha_{\mathrm{min}}$ from NGC, SGC, and the total (which is NGC+SGC) from the \texttt{Abacus altMTL} mocks. The solid black vertical line corresponds to the actual measurement of the BAO scale from the DESI DR1 3PCF. The legends of this panel show the mean of $\alpha_{\mathrm{min}}$ and the uncertainty of the mean, which is the standard deviation of the distribution divided by the square root of the number of mocks and is denoted as (stat.). The offset of $\alpha_{\rm min}$ from $1$ shows the systematic error on $\alpha$. The left panel indicates whether the measurement of the BAO scale itself (the $\alpha_{\mathrm{min}}$ we measured) is consistent with the mocks.

The right panel of \Cref{fig:alpha_EZmock} shows the precision on the BAO scale (which is denoted as $\sigma_{\alpha}$) from  mocks compared to the data. We have obtained the precision of the BAO scale from mocks in the same way we did for the data itself, as was explained in \cref{sec:BAO detection Sig.}. This panel shows whether the precision on the BAO scale is consistent with the mocks. We can see that the $\Delta \alpha$ from the data (black solid vertical line) is slightly larger than the $\sigma_\alpha$ measured from the mocks (green histogram which is for the NGC+SGC). This plot also shows that the statistical error obtained from the data matches that of the mocks, which is expected. We note that the precision on the BAO scale should not be confused with the statistical error on $\alpha_{\rm min}$ which is denoted by (stat.) in the left panels of \Cref{fig:alpha_EZmock}.

We can obtain the systematic error due to our modeling from the left panel of \Cref{fig:alpha_EZmock}. The mocks allow us to estimate the systematic error and add it to the statistical error in quadrature, to find the total error budget of our measurements of the BAO scale. Doing so, we find:
\begin{empheq}[box=\fbox]{align}
{\rm \texttt{EZmock}\; Covariance}:\;\alpha = \num{0.991 +- 0.017}\;(\mathrm{total}), 
\label{eq:alpha_zfulsampleEz_tot} \\
{\rm Analytical\; Covariance}:\;\alpha = \num{0.981 +- 0.011}\;(\mathrm{total}),
\label{eq:alpha_zfulsampleAn_tot}
\end{empheq}
which is \SI{1.72}{\percent} precision on the BAO scale using the \texttt{EZmock} covariance and \SI{1.12}{\percent} for the analytical covariance. The $D_{\mathrm{V}}/r_{\mathrm{d}}$ for this sample, given the total error (systematic + statistical) is then given by:
\begin{empheq}[box=\fbox]{align}
{\rm \texttt{EZmock}\; Covariance}:\;D_{\mathrm{V}}(z = 0.68)/r_{\mathrm{d}} = \num{15.88 +- 0.27}\;(\mathrm{total}),\\
{\rm Analytical\; Covariance}:\;D_{\mathrm{V}}(z = 0.68)/r_{\mathrm{d}} = \num{15.72 +- 0.18}\;(\mathrm{total}).
\end{empheq}

\subsection{Summary of the Measurements}

\begin{figure}[h]
    \centering
    \includegraphics[width=0.6\textwidth]{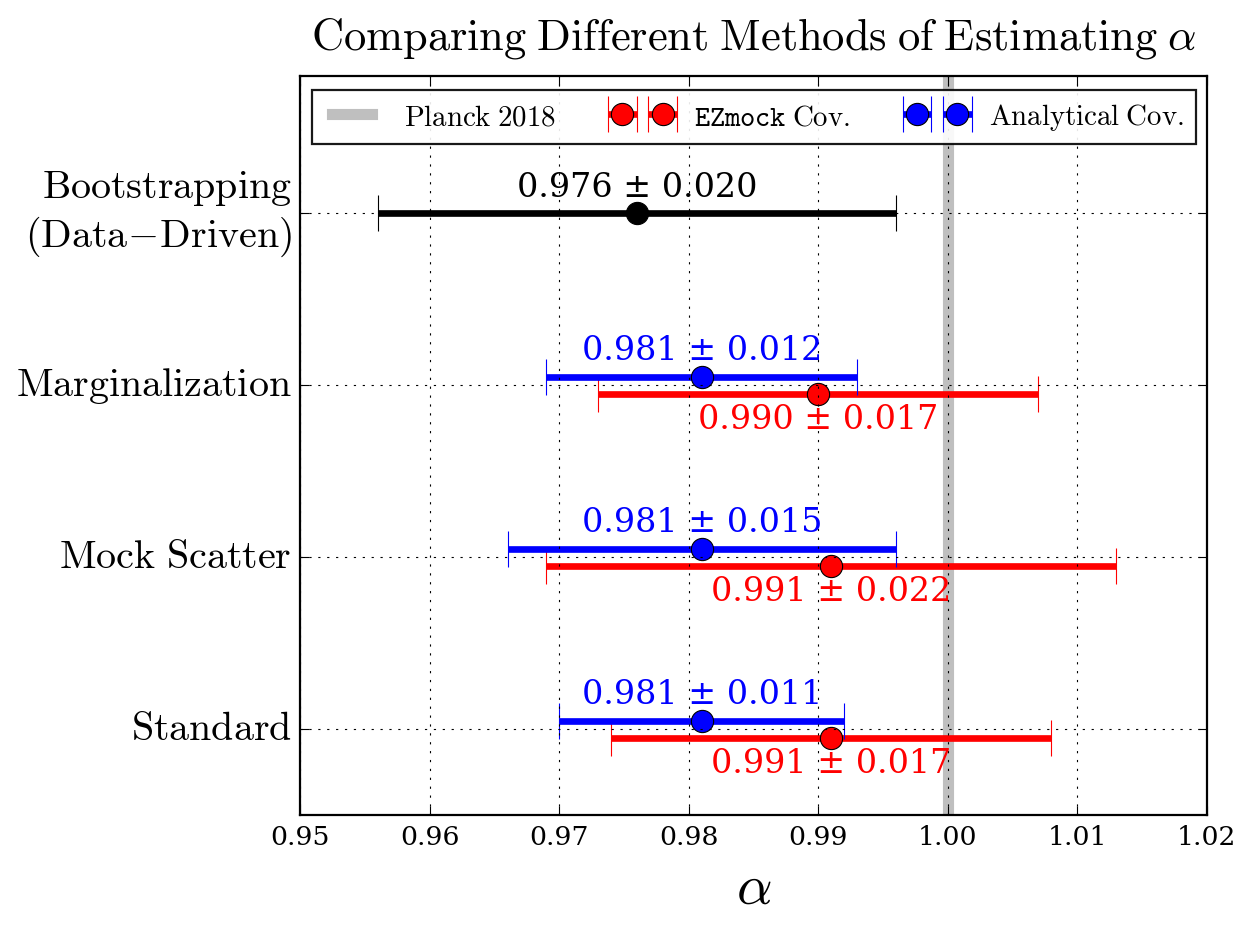}

    \caption{We measured $\alpha$ using several different methods in this work and summarize the results in this figure. The plot shows that all measurements are consistent with one another and also agree with the \textit{Planck} 2018 cosmology \cite{Planck:2018vyg} at the $2\sigma$ level. The standard method, discussed in \S\ref{sec:BAO detection Sig.}, constitutes the main result of this work. The bootstrapping, mock scatter, and marginalization methods are presented in Appendices \S\ref{sec:Bootstrap}, \S\ref{sec:MocksMethods}, and \S\ref{sec:MargMethod}, respectively. The bootstrapping method is entirely data-driven and does not depend on the choice of covariance matrix for estimating error bars; therefore, we represent it with a single black error bar. Similarly, the error bars from the mock scatter method are independent of the covariance matrix. This plot demonstrates that the estimation of both the BAO scale and its precision is consistent across different methods.}
    \label{fig:summary}
\end{figure}

Let us summarize the results obtained from the different methods used to estimate $\alpha$ and its precision. To recap, we employ four different and independent approaches to measure the BAO scale in this work. The standard method is discussed in the main text in \S\ref{sec:BAO detection Sig.}, while the other three methods are presented in the appendices, \S\ref{sec:Bootstrap}, \S\ref{sec:MocksMethods} and \S\ref{sec:MargMethod}. A summary of the results is shown in \Cref{fig:summary}.

All of these measurements are consistent with the \textit{Planck} 2018 cosmology \cite{Planck:2018vyg} at the $2\sigma$ level and are also highly consistent with one another. We further find that all measurements prefer a BAO scale that is marginally smaller than the prediction from \textit{Planck} 2018 \cite{Planck:2018vyg}, in agreement with DESI DR1 post-reconstructed 2PCF BAO measurements \cite{DESI:2024aax, DESI:2024lzq, DESI:2024mwx, DESI:2024uvr}.

\section{Comparison with DESI DR1 2PCF Measurements}

\begin{figure}[h]
    \centering
    \includegraphics[width=0.6\textwidth]{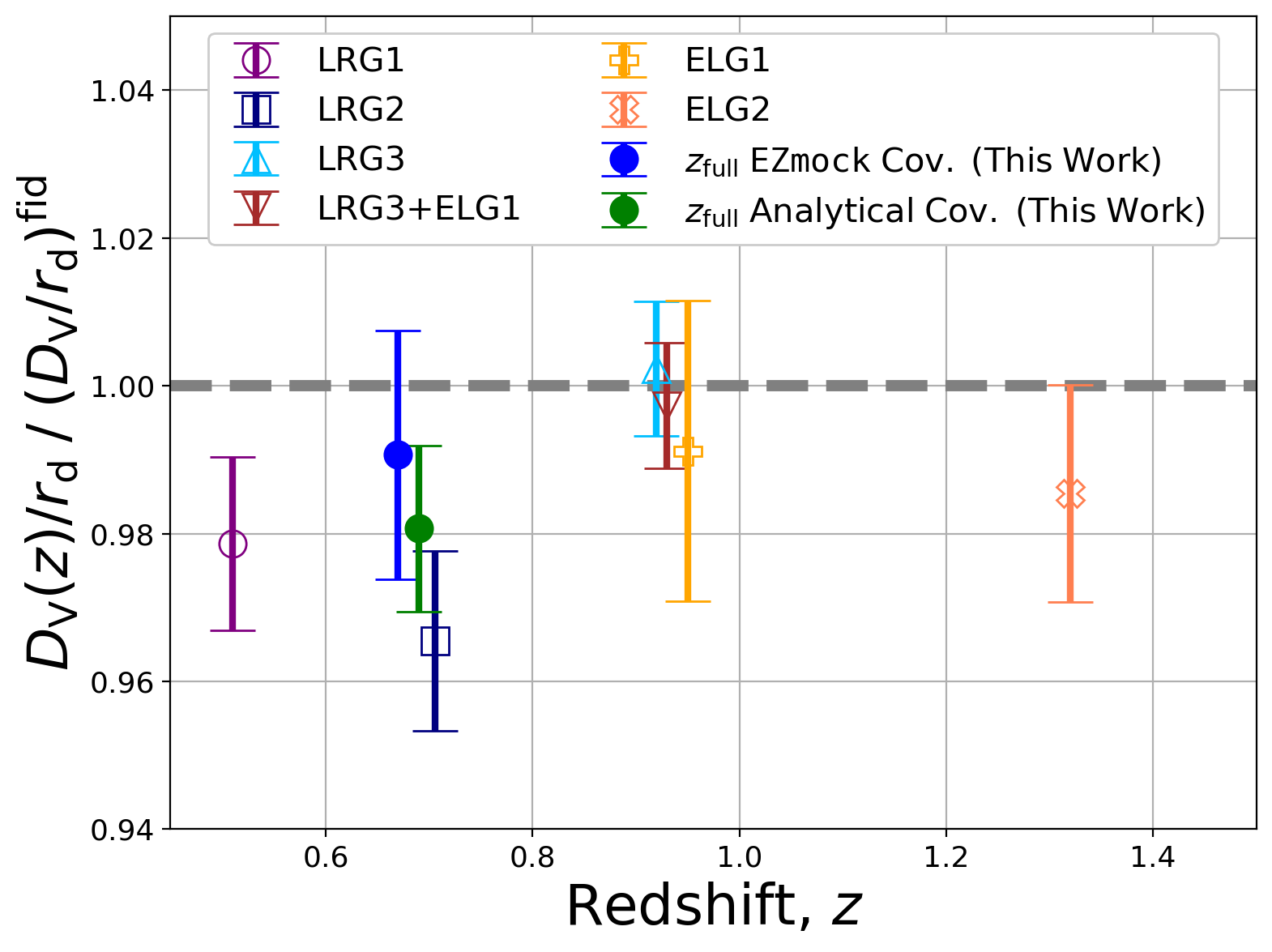}

    \caption{The ratio of the angle-averaged distance to the sound horizon at the drag epoch to its fiducial value, \( (D_{\mathrm{V}}/r_{\mathrm{d}})/(D_{\mathrm{V}}/r_{\mathrm{d}})^{\rm fid} \), is plotted as a function of redshift. The dashed gray line shows the prediction from the fiducial model, which assumes a flat \( \Lambda \)CDM cosmology with parameters described in \cref{sec:W-vs-NoW}. Although the \( z_{\mathrm{full}} \) results (blue and green points) correspond to the same redshift, we have slightly shifted them for visual clarity. For comparison, we also show measurements of the same quantity from the DESI 2PCF analysis~\cite{DESI:2024mwx, DESI:2024uvr}, derived from various tracers (LRGs and ELGs) at different redshifts. We note again that our analysis is performed over the full LRG redshift range (LRG1 + LRG2 + LRG3). The combined BAO constraint from the three LRG samples is $0.6\%$, which is 2-3 times tighter than the 3PCF-only constraint, depending on the choice of the covariance matrix. As shown, the 3PCF measurements are more consistent with the predictions of \textit{Planck}~2018~\cite{Planck:2018vyg}. Moreover, the uncertainty on the BAO scale from the post-reconstructed 2PCF is smaller than that of the 3PCF measurements at the same redshift~\cite{DESI:2024mwx, DESI:2024uvr}, which is expected.}
    \label{Fig:Dv}
\end{figure}

\Cref{Fig:Dv} shows the measurement presented in this paper, which is solely the BAO measurement from the pre-reconstructed 3PCF of the DESI DR1 $z_{\rm full}$ sample (LRG1 + LRG2 + LRG3), compared to the previous measurement from the post-reconstructed DESI DR1 2PCF \cite{DESI:2024mwx, DESI:2024uvr}. As evident from the plot, the error bars on $D_{\rm {V}}/r_{\mathrm{d}}$ obtained from the 3PCF are comparable to other DESI tracers at several different redshifts. We are comparing our measurement of the BAO scale to the measurement of the BAO scale from the LRG1 (effective redshift $z_{\mathrm{eff}} = 0.51$), LRG2 ($z_{\mathrm{eff}} = 0.70$), LRG3 ($z_{\mathrm{eff}} = 0.92$), LRG3+ELG1 ($z_{\mathrm{eff}} = 0.93$), ELG1 ($z_{\mathrm{eff}} = 0.95$) and the ELG2 ($z_{\mathrm{eff}} = 1.32$) samples. \footnote{The information presented here about the different samples and their effective redshifts is taken from Table 1 of \cite{DESI:2024mwx} and Table 18 of \cite{DESI:2024uvr}.} The uncertainty on the BAO scale from the 3PCF is larger than that from the post-reconstructed 2PCF. For example, the combined BAO scale measurement in the LRG sample (LRG1 + LRG2 + LRG3) is $\alpha = 0.987 \pm 0.006$, corresponding to a $0.6\%$ precision \cite{DESI:2024uvr}. In comparison, for the pre-reconstructed 2PCF, DESI reports $\alpha = 0.988 \pm 0.009$, which corresponds to $0.8\%$ precision (see Table 15 of \cite{DESI:2024uvr}). One reason for the lower precision of the 3PCF compared to the 2PCF is that we used only a single redshift bin for our measurement, whereas the 2PCF measurements are divided into three non-overlapping redshift bins. Splitting the data into multiple bins may increase the precision on the BAO scale. We defer the analysis using three redshift bins to future work.

We can also compare our BAO detection significance to that obtained from the reconstructed 2PCF, which shows a detection at $12.8\sigma$~\cite{DESI:2024uvr}. In contrast, we find $8.1\sigma$ with the \texttt{EZmock} covariance and $8.5\sigma$ with the analytical covariance, demonstrating that the post-reconstructed 2PCF outperforms the pre-reconstructed 3PCF in detecting the BAO feature.

We also notice that our measurement of the BAO scale is in less tension with the \textit{Planck} 2018 results (which represent our fiducial cosmology) than the BAO scale measurement of \cite{DESI:2024mwx, DESI:2024uvr} since the error bars on our measurements are higher than the 2PCF. At roughly the same redshift ($z \sim 0.70$), the 2PCF LRG2 measurement appears to be in about $2\sigma$ tension with \textit{Planck} 2018, whereas the 3PCF measurement lies well within the $1\sigma$ range of the fiducial prediction (blue points in \Cref{Fig:Dv}). Our 3PCF results show excellent agreement with the previous 2PCF measurements. According to \cref{Fig:Dv}, using the analytical covariance matrix would place our results in slightly more tension with the \textit{Planck} 2018 cosmology \cite{Planck:2018vyg}. However, this tension disappears at the $2\sigma$ level.

\section{Discussion \& Conclusions}

In this paper, we detect the BAO signal in the 3PCF of the DESI DR1 LRG sample, which is the largest LRG sample to date. We find an $8.5\sigma$ (using the analytical covariance matrix) and a $8.1\sigma$ (using the \texttt{EZmock} covariance matrix) detection of the BAO feature in the 3PCF of the DESI DR1 for the $z_{\mathrm{full}}$ sample ($0.4 \leq z \leq 1.1$). We find a stronger 3PCF BAO feature in the DESI DR1 LRGs compared to the SDSS BOSS CMASS sample, which reported a $4.5\sigma$ detection significance~\cite{SE_Full3PCF_BAO} (see also the $4.1\sigma$ BAO detection obtained from the bispectrum analysis of the same dataset; \cite{Pearson:2017wtw}).

It is known that the 3PCF is statistically not independent from the reconstructed 2PCF and the reconstruction pushes the BAO information from the 3PCF back to the 2PCF. However, the 3PCF is an independent method that may have different systematics form the 2PCF so it is still worth pursuing \cite{Schmittfull:2015mja, Samushia:2021ixs}.

We compare our measurement of the BAO scale with the \texttt{Abacus altMTL} mocks, which are the most accurate mocks currently available for the DESI DR1 analysis, designed to closely reproduce the DESI fiber assignment. Performing the analysis on the mocks allows us to estimate the systematic error. We find this systematic error to be only \SI{0.6}{\percent}. We also find no statistically significant difference between the data and the mocks. 
We have used several methods to estimate the uncertainties on the BAO scale in this work. Our first method is to take the width of the $\chi^2_{\rm min}+1$, which provides the standard result of this paper, yielding $1.72\%$ (\texttt{Ezmock}) and $1.12\%$ (analytical) precision on $\alpha$. This marks the most precise measurement of the BAO scale using the galaxy 3PCF to date. For comparison, \cite{SE_Full3PCF_BAO} reported a \SI{1.7}{\percent} precision using the SDSS BOSS CMASS DR12 sample with the analytical covariance matrix, whereas our measurement achieves $1.12\%$ precision with the analytical covariance.

The second method (discussed in more detail in \S\ref{sec:MocksMethods}) uses the width of the distribution of $\alpha_{\rm min}$ across the \texttt{Abacus altMTL} mocks, yielding $2.23\%$ (\texttt{Ezmock}) and $1.53\%$ (analytical). Another approach (discussed in \S\ref{sec:MargMethod}) marginalizes $\alpha$ over the $\chi^2$ distribution, resulting in $1.72\%$ (\texttt{Ezmock}) and $1.22\%$ (analytical). The results from this method are very similar to the standard method presented in \S\ref{sec:comparison}. The final method applies bootstrap sampling on $\alpha$ from our seven patches (described in \S\ref{sec:Bootstrap}), yielding a $2.05\%$ precision. This approach provides a covariance-free estimate of $\alpha$ since it is entirely data-driven. All of these results are consistent with each other, despite being obtained using different methods.

Additionally, we find that the BAO scale measured from the 3PCF of LRGs is in complete agreement with the previous 2PCF measurements of DESI DR1 for both LRGs and ELGs at the same redshift (\cref{Fig:Dv}).

This work demonstrates that the 3PCF of galaxies can substantially enhance our understanding of the Universe. Moreover, the 3PCF offers a complementary avenue for probing structure formation, as it is methodologically independent of the 2PCF.

Our implementation of the 3PCF is based on the FFTLog algorithm (which is very fast), which offers several advantages over previous analytical approaches such as \cite{SE_3PCF_model}. First, it allows the inclusion of higher-order loop corrections to the tree-level bispectrum, enabling the model to more accurately describe the data on smaller scales. Second, it can be integrated with the 2PCF pipeline, paving the way for a future full-shape joint 2PCF+3PCF analysis.

\begin{acknowledgments}

FK, ZS, AG acknowledge support from NASA grant number 80NSSC24M0021. ZS thanks D. Eisenstein, D.\ Slepian, M.\ Slepian, I.\ Slepian, A.\ Miller for useful conversations. WOL acknowledges support from Grant 63041 from the John Templeton Foundation.
AK was supported as a CITA National Fellow by the Natural Sciences and Engineering Research Council of Canada (NSERC), funding reference \#DIS-2022-568580. 

Research at the Perimeter Institute is supported in part by the Government of Canada through the Department of Innovation, Science and Economic Development Canada and by the Province of Ontario through the Ministry of Colleges and Universities.

This material is based upon work supported by the U.S. Department of Energy (DOE), Office of Science, Office of High-Energy Physics, under Contract No. DE–AC02–05CH11231, and by the National Energy Research Scientific Computing Center, a DOE Office of Science User Facility under the same contract. Additional support for DESI was provided by the U.S. National Science Foundation (NSF), Division of Astronomical Sciences under Contract No. AST-0950945 to the NSF’s National Optical-Infrared Astronomy Research Laboratory; the Science and Technology Facilities Council of the United Kingdom; the Gordon and Betty Moore Foundation; the Heising-Simons Foundation; the French Alternative Energies and Atomic Energy Commission (CEA); the National Council of Humanities, Science and Technology of Mexico (CONAHCYT); the Ministry of Science, Innovation and Universities of Spain (MICIU/AEI/10.13039/501100011033), and by the DESI Member Institutions: \url{https://www.desi.lbl.gov/collaborating-institutions}. Any opinions, findings, and conclusions or recommendations expressed in this material are those of the author(s) and do not necessarily reflect the views of the U. S. National Science Foundation, the U. S. Department of Energy, or any of the listed funding agencies.

The authors are honored to be permitted to conduct scientific research on I'oligam Du'ag (Kitt Peak), a mountain with particular significance to the Tohono O’odham Nation.
\end{acknowledgments}

\appendix

\section{Bootstrap Re-Sampling on \texorpdfstring{$\alpha$}{alpha} from the \texorpdfstring{$z_{\rm full}$}{zfull} Sample Patches}
\label{sec:Bootstrap}
Another method to estimate the errorbar on the BAO scale is to re-sample the $\alpha$ measured from the $z_{\rm full}$ sample (which spans the redshift range of $0.4<z<1.1$) using the bootstrapping method. Within the $z_{\mathrm{full}}$ sample, we divide the North Galactic Cap (NGC) into four patches and the South Galactic Cap (SGC) into three patches, resulting in a total of seven spatially distinct and well-separated patches. These patches are selected such that the sample will have high completeness. The ranges in right ascension (RA) and declination (Dec) for each of these patches and regions are listed in \cref{tab:Regions}. The smaller volumes in each patch degrade the signal-to-noise ratio (SNR).  
\begin{table}[h]
    \centering

    \caption{Right ascension (RA) and Declination (Dec) of the patches in the $z_{\rm full}$ sample used in our bootstrapping analysis. The $z_{\mathrm{full}}$ sample includes all galaxies in the NGC and SGC up to $z = 1.1$. In the second row, we report the measured BAO scale from each individual patch. The volume of the patches are very similar to each other, but we account for the small differences in volume in our bootstrapping analysis as well.}
    \smallskip

    \setlength{\tabcolsep}{4pt}
    \begin{tabular}{l l cccc c ccc}
        \toprule
        & &
        \multicolumn{4}{c}{\textbf{NGC}} &
        &
        \multicolumn{3}{c}{\textbf{SGC}}
        \\
        \cmidrule(lr){3-6} \cmidrule(lr){8-10}
        \textbf{\boldmath $z_{\mathrm{full}}$ Patches} &
        \makecell[l]{RA: \\ Dec:} &
        \makecell[l]{$[180, 260]^{\circ}$ \\ $[30, 40]^{\circ}$} &
        \makecell[l]{$[110, 160]^{\circ}$ \\ $[-10, 8]^{\circ}$} &
        \makecell[l]{$[170, 200]^{\circ}$ \\ $[-10, 8]^{\circ}$} &
        \makecell[l]{$[210, 260]^{\circ}$ \\ $[-10, 8]^{\circ}$} &
        &
        \makecell[l]{$[35, 80]^{\circ}$   \\ $[-20, 30]^{\circ}$} &
        \makecell[l]{$[0-25]^{\circ}$     \\ $[-20,25]^{\circ}$} &
        \makecell[l]{$[300, 350]^{\circ}$ \\ $[-20,25]^{\circ}$}
        \\
        \midrule
        \textbf{\boldmath Measured $\alpha$} &
        \makecell[l]{} &
        \makecell[l]{0.924} & \makecell[l]{1.000} &
        \makecell[l]{0.891} & \makecell[l]{0.994} &
        & \makecell[l]{0.891} & \makecell[l]{1.009} & \makecell[l]{1.102}
        \\
        \bottomrule
    \end{tabular}
    \label{tab:Regions}
\end{table}
These patches are explained in more detail in \cite{Krolewski:2024paz, hou2025study, slepian2025measurement}. We find that the spread of $\alpha$ is consistent with the statistical uncertainty on $\alpha$ measured from the \texttt{Abacus altMTL} mocks, as shown in \Cref{fig:alpha_EZmock}.

We resample the BAO scale from these patches 1000 times and use the resulting distribution to estimate the mean and uncertainty of the BAO scale in the $z_{\rm full}$ sample. Although the volumes of the patches are very similar to each other, we account for their small differences when performing the resampling. Since the total volume of the patches corresponds to only $58\%$ of the total $z_{\rm full}$ sample volume, we rescale the uncertainty on $\alpha$ to account for the reduced volume.\footnote{This correction factor is obtained by comparing the number of randoms used in the measurement of the 3PCF.} After applying this correction, we obtain the best–fit value of $\alpha$ together with its rescaled uncertainty.:
\begin{align}
    \alpha = 0.976 \pm 0.020 .
    \label{Eq:Bootstrap}
\end{align}
This corresponds to a $2.05\%$ precision on the BAO scale and thus, is the most conservative upper bound. This procedure provides a fully data-driven estimate of the BAO uncertainty without requiring mock catalogs.

We could also, in principle, determine the mean and standard deviation of $\alpha$ across all seven patches and use that distribution to measure the BAO scale (without the bootstrap sampling). In this case, since the measurement of $\alpha$ is performed on a single patch, we must rescale the uncertainty by the square root of the ratio between the mean patch volume and the total sample volume, which is $0.29$. Using this approach yields the same result as \cref{Eq:Bootstrap}, as expected.

We have also performed this analysis on a single \texttt{Ezmock} realization (since the replication factor does not need to be applied to them \cite{slepian2025measurement}). The value of $\alpha$ estimated from this realization (from seven patches) using the bootstrapping method is $\alpha = 0.990 \pm 0.006$. We find that the spread of $\alpha$ across patches in the \texttt{Ezmock} realization is smaller than in the data, indicating that the uncertainty obtained here, which is entirely derived from the data, is very conservative.

\section{Mock Scatter Method}
\label{sec:MocksMethods}

Another upper bound on $\alpha$, which we consider to be very conservative, can be obtained from the distribution of $\alpha_{\rm min}$ in the mocks, as shown in the left panels of \Cref{fig:alpha_EZmock}. This conservative bound is obtained by replacing the statistical error bar on $\alpha$ from the data with the $1\sigma$ width of the distribution of $\alpha_{\rm min}$ in the mocks. We then add the systematic offset to the statistical error bar as before to find:
\begin{align}
    &{\rm \texttt{EZmock}\; Covariance}:\;\alpha = \num{0.991 +- 0.022}\;(\mathrm{total})\;
    \label{eq:alpha_zfulsampleEz_tot2},\\
    &{\rm Analytical\; Covariance}:\;\alpha = \num{0.981 +- 0.015}\;(\mathrm{total})\;,
    \label{eq:alpha_zfulsampleAn_tot2}
\end{align}
which correspond to $2.23\%$ precision from \texttt{Ezmocks} and $1.53\%$ from analytical covariance matrix.

\section{Marginalization of \texorpdfstring{$\chi^2$}{chi2} Over $\alpha$}
\label{sec:MargMethod}

\begin{figure}[h]
    \centering
    \includegraphics[width=0.98\textwidth]{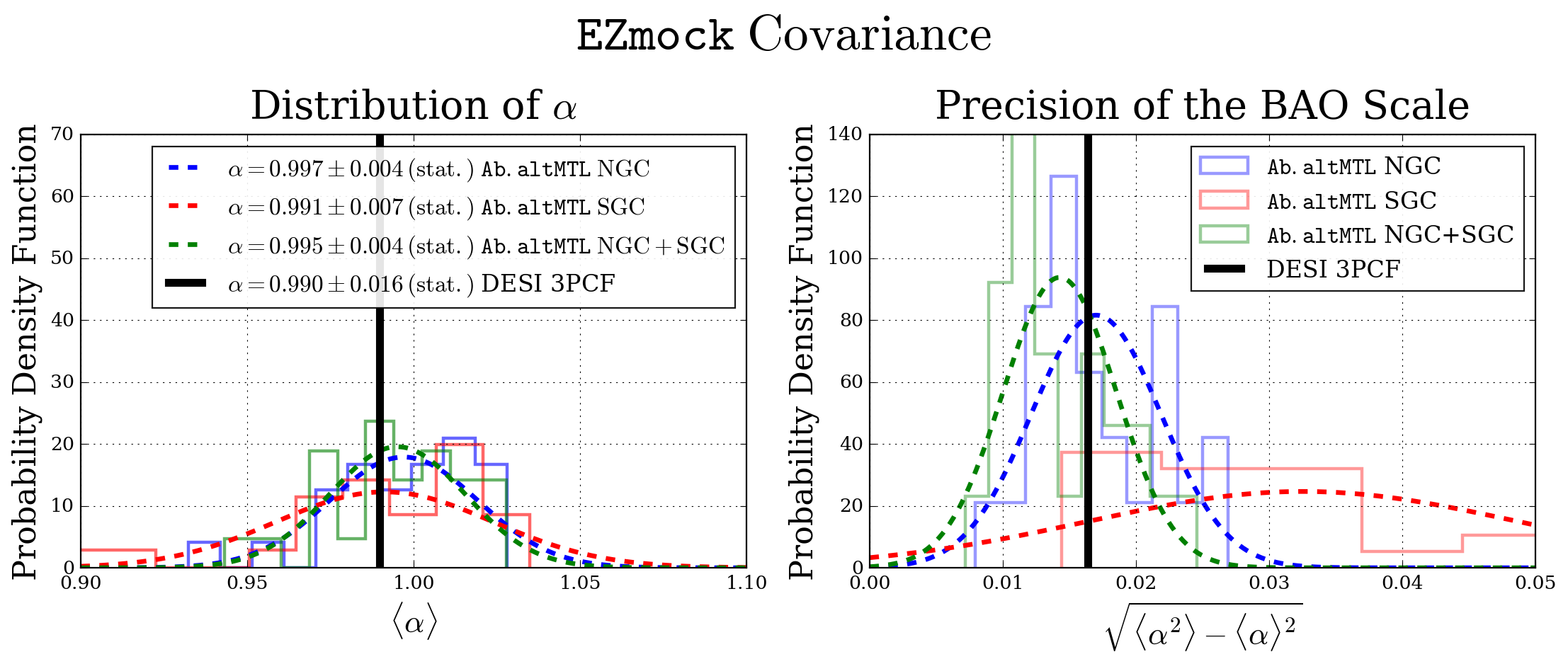}
    \vspace{0.3cm}
    \includegraphics[width=0.98\textwidth]{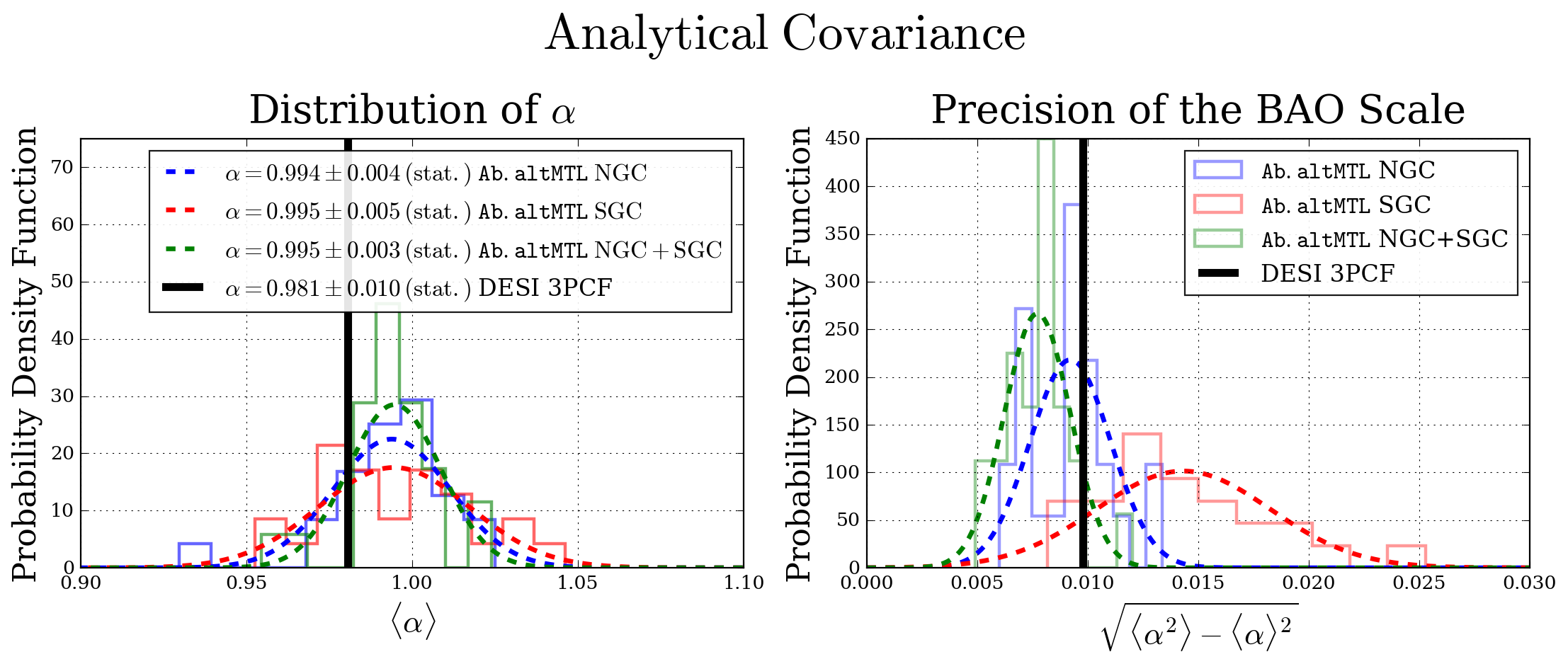}
    \caption{\textit{Left:} 
    Same as \Cref{fig:alpha_EZmock} but integrating $\alpha$ against the likelihood given by \cref{eq:alpha_another_method} to obtain the mean and error bar on the BAO scale. Our constraints on $\alpha$ using this method is consistent with our standard measurement (\Cref{eq:alpha_zfulsampleEz_tot} and \Cref{eq:alpha_zfulsampleAn_tot}). As was also seen from \Cref{fig:alpha_EZmock}, we see very good agreement between the data and the mocks.}
    \label{fig:anothermethod}
\end{figure}

Another method to estimate the mean and error bar of the BAO scale is to marginalize $\alpha$ against the $\chi^2$ distribution. We use
\begin{align}
\langle \alpha^n \rangle = \frac{1}{N_\alpha} \int d\alpha \, \alpha^n \, \exp{\left[-\frac{\chi^2(\alpha)}{2}\right]},
\label{Eq:Alpha_another}
\end{align}
to estimate the mean and variance of $\alpha$ for both the mocks and the data. Here, $N_\alpha$ in Eq.~\ref{Eq:Alpha_another} is the normalization factor and is equal to the same integral evaluated at $n=0$. We evaluate \cref{Eq:Alpha_another} for \(n = 0, 1, 2\). The case \(n = 0\) gives the normalization factor, \(n = 1\) provides the mean value of \(\alpha\), and \(n = 2\) allows us to compute the uncertainty on \(\alpha\) via
\(\sigma_{\alpha} = \sqrt{\langle \alpha^2 \rangle - \langle \alpha \rangle^2}\). This method was previously used to estimate the error bars on the BAO scale from the BOSS CMASS DR12 sample \cite{SE_Full3PCF_BAO} and provides a separate, independent measure of the BAO scale and precision. Using the $\chi^2$ distribution we obtained on the mocks and the data we measure the BAO scale as:
\begin{align}
    &{\rm \texttt{EZmock}\; Covariance}:\;\alpha = \num{0.990 +- 0.017}\;(\mathrm{total})\;
    \label{eq:alpha_another_methodEZmocks},\\
    &{\rm Analytical\; Covariance}:\;\alpha = \num{0.981 +- 0.012}\;(\mathrm{total})\;,
    \label{eq:alpha_another_method}
\end{align}
which is $1.72\%$ precision from \texttt{EZmocks} and $1.22\%$ precision from the analytical covariance matrix. These constraints with the marginalization method is completely consistent with the standard method we report in \Cref{eq:alpha_zfulsampleEz_tot} and \Cref{eq:alpha_zfulsampleAn_tot}. \Cref{fig:anothermethod} shows the same plot as \Cref{fig:alpha_EZmock}, but using the marginalization approach. 

\section{Analytical Covariance Rescaling}
\label{sec:Rescale}

From \Cref{Fig:zfulltotal}, we see that the $\chi^2$/d.o.f for the analytical covariance matrix is $590.2 / 521$ for the NGC and $735.6 / 521$ for the SGC. We can multiply the analytical covariance matrix by a constant to bring the $\chi^2$ and the dof into exact agreement. These rescaling constants are $0.88$ for the NGC and $0.71$ for the SGC. Multiplying the analytical covariance by these factors is equivalent to reducing the volumes reported in \S\ref{subsec:Measurement} by the same constants.

We note that rescaling the covariance matrix does not change the location of the minima in the $\chi^2$ distribution for a single sample. However, because the rescaling coefficients differ for the NGC and SGC, their combination leads to a slightly different minimum in the overall $\chi^2$ distribution. As a result, both the overall BAO location and the systematic offsets are affected. In practice, however, the systematic offset changes only marginally.

As a result of this rescaling, both the BAO detection significance and the precision of the BAO scale decrease. After rescaling, the BAO signal in the 3PCF can be detected at $7.7\sigma$, and the BAO scale is measured at $\alpha = 0.978 \pm 0.012$ including the systematic error, corresponding to a $1.24\%$ precision (the standard method achieves $1.12\%$ precision). This approach is more conservative than our standard method with the analytical covariance matrix, but still less conservative than using the \texttt{Ezmock} covariance matrix (with $1.72\%$ precision). We note that since the \texttt{Ezmock} covariance is already multiplied by both the Hartlap and Percival factors, we do not rescale the covariance further. 

\section{Author affiliations}
\label{AppE:Authors}
\noindent 
The affiliations of the authors are listed in \S\ref{AppE:Authors}.\\
\textsuperscript{1}Department of Physics, University of Florida, 2001 Museum Rd., Gainesville, FL 32611, USA\\
\textsuperscript{2}Department of Astronomy, University of Florida, 211 Bryant Space Science Center, Gainesville, FL 32611, USA\\
\textsuperscript{3}Waterloo Centre for Astrophysics, Department of Physics and Astronomy, University of Waterloo, 200 University Avenue West, Waterloo, ON N2L 3G1, Canada\\
\textsuperscript{4}Department of Physics and Astronomy, University of Waterloo, Waterloo, ON N2L 3G1, Canada\\
\textsuperscript{5}Instituto de Astrof\'{i}sica de Andaluc\'{i}a (CSIC), Glorieta de la Astronom\'{i}a, s/n, E-18008 Granada, Spain \\
\textsuperscript{6}Lawrence Berkeley National Laboratory, 1 Cyclotron Road, Berkeley, CA 94720, USA \\
\textsuperscript{7}Department of Physics, Boston University, 590 Commonwealth Avenue, Boston, MA 02215 USA \\
\textsuperscript{8} Dipartimento di Fisica ``Aldo Pontremoli'', Universit\`a degli Studi di Milano, Via Celoria 16, I-20133 Milano, Italy \\
\textsuperscript{9} INAF-Osservatorio Astronomico di Brera, Via Brera 28, 20122 Milano, Italy \\
\textsuperscript{10}Department of Physics \& Astronomy, University College London, Gower Street, London, WC1E 6BT, UK \\
\textsuperscript{11}Department of Physics and Astronomy, The University of Utah, 115 South 1400 East, Salt Lake City, UT 84112, USA \\
\textsuperscript{12}Instituto de F\'{\i}sica, Universidad Nacional Aut\'{o}noma de M\'{e}xico,  Circuito de la Investigaci\'{o}n Cient\'{\i}fica, Ciudad Universitaria, Cd. de M\'{e}xico  C.~P.~04510,  M\'{e}xico \\
\textsuperscript{13}Department of Physics \& Astronomy, University of Rochester, 206 Bausch and Lomb Hall, P.O. Box 270171, Rochester, NY 14627-0171, USA \\
\textsuperscript{14}Space Sciences Laboratory, University of California, Berkeley, 7 Gauss Way, Berkeley, CA  94720, USA \\
\textsuperscript{15}University of California, Berkeley, 110 Sproul Hall \#5800 Berkeley, CA 94720, USA \\
\textsuperscript{16}Departamento de F\'isica, Universidad de los Andes, Cra. 1 No. 18A-10, Edificio Ip, CP 111711, Bogot\'a, Colombia \\
\textsuperscript{17}Observatorio Astron\'omico, Universidad de los Andes, Cra. 1 No. 18A-10, Edificio H, CP 111711 Bogot\'a, Colombia \\
\textsuperscript{18}Institut d'Estudis Espacials de Catalunya (IEEC), c/ Esteve Terradas 1, Edifici RDIT, Campus PMT-UPC, 08860 Castelldefels, Spain \\
\textsuperscript{19}Institute of Cosmology and Gravitation, University of Portsmouth, Dennis Sciama Building, Portsmouth, PO1 3FX, UK \\
\textsuperscript{20}Institute of Space Sciences, ICE-CSIC, Campus UAB, Carrer de Can Magrans s/n, 08913 Bellaterra, Barcelona, Spain \\
\textsuperscript{21}University of Virginia, Department of Astronomy, Charlottesville, VA 22904, USA \\
\textsuperscript{22}Fermi National Accelerator Laboratory, PO Box 500, Batavia, IL 60510, USA \\
\textsuperscript{23}Institut d'Astrophysique de Paris. 98 bis boulevard Arago. 75014 Paris, France \\
\textsuperscript{24}IRFU, CEA, Universit\'{e} Paris-Saclay, F-91191 Gif-sur-Yvette, France \\
\textsuperscript{25}Center for Cosmology and AstroParticle Physics, The Ohio State University, 191 West Woodruff Avenue, Columbus, OH 43210, USA \\
\textsuperscript{26}Department of Physics, The Ohio State University, 191 West Woodruff Avenue, Columbus, OH 43210, USA \\
\textsuperscript{27}The Ohio State University, Columbus, 43210 OH, USA \\
\textsuperscript{28}School of Mathematics and Physics, University of Queensland, Brisbane, QLD 4072, Australia \\
\textsuperscript{29}Department of Physics, University of Michigan, 450 Church Street, Ann Arbor, MI 48109, USA \\
\textsuperscript{30}University of Michigan, 500 S. State Street, Ann Arbor, MI 48109, USA \\
\textsuperscript{31}Department of Physics, The University of Texas at Dallas, 800 W. Campbell Rd., Richardson, TX 75080, USA \\
\textsuperscript{32}NSF NOIRLab, 950 N. Cherry Ave., Tucson, AZ 85719, USA \\
\textsuperscript{33}Department of Physics and Astronomy, University of California, Irvine, 92697, USA \\
\textsuperscript{34}Sorbonne Universit\'{e}, CNRS/IN2P3, Laboratoire de Physique Nucl\'{e}aire et de Hautes Energies (LPNHE), FR-75005 Paris, France \\
\textsuperscript{35}Departament de F\'{i}sica, Serra H\'{u}nter, Universitat Aut\`{o}noma de Barcelona, 08193 Bellaterra (Barcelona), Spain \\
\textsuperscript{36}Institut de F\'{i}sica d’Altes Energies (IFAE), The Barcelona Institute of Science and Technology, Edifici Cn, Campus UAB, 08193, Bellaterra (Barcelona), Spain \\
\textsuperscript{37}Instituci\'{o} Catalana de Recerca i Estudis Avan\c{c}ats, Passeig de Llu\'{\i}s Companys, 23, 08010 Barcelona, Spain \\
\textsuperscript{38}Department of Physics and Astronomy, Siena University, 515 Loudon Road, Loudonville, NY 12211, USA \\
\textsuperscript{39}Department of Physics \& Astronomy and Pittsburgh Particle Physics, Astrophysics, and Cosmology Center (PITT PACC), University of Pittsburgh, 3941 O'Hara Street, Pittsburgh, PA 15260, USA \\
\textsuperscript{40}Perimeter Institute for Theoretical Physics, 31 Caroline St. North, Waterloo, ON N2L 2Y5, Canada \\
\textsuperscript{41}Departament de F\'isica, EEBE, Universitat Polit\`ecnica de Catalunya, c/Eduard Maristany 10, 08930 Barcelona, Spain \\
\textsuperscript{42}Abastumani Astrophysical Observatory, Tbilisi, GE-0179, Georgia \\
\textsuperscript{43}Department of Physics, Kansas State University, 116 Cardwell Hall, Manhattan, KS 66506, USA \\
\textsuperscript{44}Faculty of Natural Sciences and Medicine, Ilia State University, 0194 Tbilisi, Georgia \\
\textsuperscript{45}CIEMAT, Avenida Complutense 40, E-28040 Madrid, Spain \\
\textsuperscript{46}Department of Physics \& Astronomy, Ohio University, 139 University Terrace, Athens, OH 45701, USA \\
\textsuperscript{47}Department of Astronomy, Tsinghua University, 30 Shuangqing Road, Haidian District, Beijing, China, 100190 \\
\textsuperscript{48}National Astronomical Observatories, Chinese Academy of Sciences, A20 Datun Road, Chaoyang District, Beijing, 100101, P.~R.~China \\

\newpage
\bibliography{apssamp}

\end{document}